\documentclass[11pt,english,longbibliography]{article}
\usepackage{url}

\usepackage{draft} 
\usepackage{putex}
\usepackage{hyperref}

\renewcommand{\Re}{{\rm Re \, }}

\newcommand{\CO}{{\cal O}}

\newcommand{\CD}{{\cal D}}

\newcommand{\CK}{{\cal K}}
\newcommand{\CW}{{\cal W}}

\newcommand{\CF}{{\cal F}}
\renewcommand{\CP}{{\cal P}}

\newcommand{\SO}{{\text{SO}}}

\usepackage{amsthm}

\makeatletter
\newcommand*{\rom}[1]{\expandafter\@slowromancap\romannumeral #1@}
\makeatother

\usepackage[latin9]{inputenc}
\usepackage{verbatim}
\usepackage{prettyref}
\usepackage[numbers,sort&compress]{natbib}
\usepackage{amsmath}
\usepackage{amssymb}
\usepackage{tikz-cd}
\tikzset{commutative diagrams/row sep/huge=4cm}
\tikzset{commutative diagrams/column sep/huge=4cm}
\tikzcdset{scale cd/.style={every label/.append style={scale=#1},
    cells={nodes={scale=#1}}}}
\usepackage{lmodern}
\usepackage{tcolorbox}
\usepackage{cases}
\usepackage{bbold}
\usepackage{bm}
\usepackage{ytableau}
\usepackage{youngtab}
\usepackage{mathtools}
\usepackage{graphicx}
\usepackage[nottoc]{tocbibind}
\usepackage{mathrsfs}
\usepackage{caption}
\usepackage{subcaption}
\usepackage{cancel}
\usepackage{float}
\usepackage{tikz}
\usetikzlibrary{arrows.meta}
\usetikzlibrary{bending}
\tikzset{
    Witten diagram/.style={
        execute at begin picture={
            \draw[blue, line width=1.5pt] circle[radius=\pgfkeysvalueof{/tikz/Witten/radius}];
            \path node (X){\phantom{X}};
        },
        baseline={(X.base)}
    },
    vertex/.style={circle,fill,inner sep=1.5pt,node contents={}},
    Witten/.cd,
    radius/.initial=3cm
}
\usetikzlibrary{patterns}
\usepackage{framed}
\usepackage{adjustbox}
\usepackage{braket}
\usepackage{slashed}
\usepackage{esvect}
\usepackage{dsfont}

\DeclarePairedDelimiter\floor{\lfloor}{\rfloor}
\usepackage{color}
\definecolor{darkgreen}{rgb}{0,0.5,0}
\definecolor{darkblue}{rgb}{0,0,0.6}
\definecolor{purple}{rgb}{0.4,.2,0.7}

\numberwithin{equation}{section}
\numberwithin{figure}{section}
\numberwithin{table}{section}

\def\CG{{\cal G}}

\def\CN{{\cal N}}
\def\CD{{\cal D}}
\def\tr{\,{\rm tr}\,}

\DeclareFontShape{OT1}{cmr}{mx}{n}{<->cmr10}{}

\newcommand{\Dn}{\gamma\cdot\nabla}
\newcommand{\lDn}{\gamma\cdot\overleftarrow\nabla}

\newcommand{\F}{{}_2F_1}

\begin{document}

\title{\centering Boundary criticality in
the Gross-Neveu-Yukawa model at higher orders
}

\authors{Oleksandr Diatlyk \worksat{\NYU}, Simone Giombi \worksat{\PUJ} and Zimo Sun \worksat{\IAS, \PUJ} }

\institution{NYU}{Center for Cosmology and Particle Physics, New York University, New York, NY 10003, USA}
\institution{PUJ}{Joseph Henry Laboratories, Princeton University, Princeton, NJ 08544, USA}
\institution{IAS}{Institute for Advanced Study, Princeton, NJ 08540, USA}

\abstract{
We extend the study of boundary criticality in the  Gross-Neveu-Yukawa universality class beyond leading order. Using the hyperbolic space formulation of boundary conformal field theories, we compute the first subleading corrections at large $N$ to the free energies of the ``normal", ``ordinary" and ``special" boundary universality classes. We also determine the order $1/N$ correction to the dimension of the boundary fermion at the normal fixed point. In the Gross-Neveu-Yukawa theory in $d=4-\epsilon$, we perform a higher-order analysis of the boundary free energy, and use it to extract estimates for the boundary central charge in $d=3$. The large $N$ and $\epsilon$-expansion results are shown to be precisely consistent in overlapping regimes, providing nontrivial consistency checks for the identification of the boundary universality classes. Our calculations rely on a combination of AdS harmonic analysis and boundary conformal field theory techniques.
}

\date{}
\maketitle
\tableofcontents

\section{Introduction}

Boundary critical phenomena provide a rich extension of the theory of bulk criticality. In the presence of a boundary, a given bulk conformal field theory may admit several distinct conformal boundary conditions, corresponding to different boundary universality classes connected by renormalization-group flows. A classic and extensively studied example is provided by the so-called ``ordinary", ``special" and ``extraordinary/normal" transitions of the scalar O$(N)$ model, see e.g. \cite{Diehl:1996kd, McAvity:1995zd, Liendo:2012hy, Carmi:2018qzm, Giombi:2020rmc,Metlitski:2020cqy}. More recently, analogous questions have been studied in theories of interacting fermions in the Gross-Neveu-Yukawa universality class, where a similar rich structure of conformal boundary phases was uncovered \cite{Giombi:2021cnr,Herzog:2022jlx, Jiang:2025sfb, Fedorenko:2026svv}. \footnote{The recent work \cite{Fedorenko:2026svv} found a richer set of conformal boundary conditions by allowing two types of fermionic boundary terms, $\bar\Psi^I\Psi_I$ and $\bar\Psi^I\gamma^5\Psi_I$. The latter can be mapped to  the pseudoscalar Yukawa model with a boundary, which was studied in \cite{Jiang:2025sfb} using the $4-\epsilon$ expansion. They found a phase structure similar to that of the Gross-Neveu-Yukawa model, but with different critical exponents. In this paper, we follow \cite{Giombi:2021cnr} and hence do not turn on $\bar\Psi^I\gamma^5\Psi_I$ on the boundary.}

The Gross-Neveu (GN) model is a theory of interacting fermions with an action
\begin{align}\label{GNaction0}
S_{\rm GN} =  - \int d^d x\, \left[\bar\Psi^I \slashed{\partial}\Psi_I+\frac{g}{4}\left(\bar\Psi^I\Psi_I\right)^2\right], \quad I=1,2,\cdots, N~.
\end{align}
Here we have set the mass of the fermions to zero, as we will be mainly interested in tuning the model to criticality. The fields $\Psi_I$ are $N$ Dirac fermions with $c_d = 2^{\floor{\frac{d}{2}}}$ complex components. The large $N$ expansion of this model can be developed by introducing a scalar auxiliary field $\sigma$ via the standard Hubbard-Stratonvich transformation, and dropping the quadratic term $\sim \sigma^2$ which becomes irrelevant in the formal UV limit. This yields the action 
\begin{align}\label{GNaction}
S_{\rm GN} =  - \int d^d x\, \bar\Psi^I \left(\slashed{\partial}+\sigma\right)\Psi_I
\end{align}
which can be used to develop the $1/N$ expansion. 

A UV completion of the GN model is provided by the  Gross-Neveu-Yukawa (GNY) model \cite{Zinn-Justin:1991ksq}
\begin{gather}\label{GNYaction}
    S_{\rm GNY}=\int d^d x \left[\frac{1}{2}\left(\partial_\mu s\right)^2- \left( \bar{\Psi}^I \slashed{\partial} \Psi_{I} + g_{1} s \bar{\Psi}^I \Psi_{I}\right) + \frac{g_{2}}{4!} s^4\right]\,.
\end{gather}
This model has IR fixed points near four dimensions which can be studied perturbatively in the framework of the Wilson-Fisher $\epsilon$-expansion. The $\beta$ functions and the IR fixed point of this model in $d=4-\epsilon$ dimensions are reviewed in Appendix \ref{betareview}. When working in continuous dimension $d=4-\epsilon$, we will consider $N$ four component Dirac fermions, and keep the number of fermion components fixed. Continuing the theory to $d=3$ and formally setting $N=1/4$ (corresponding to a single Majorana fermion) yields a theory with ``emergent" ${\cal N}=1$ supersymmetry \cite{Iliesiu:2015qra, Grover:2013rc, Bashkirov:2013vya, Fei:2016sgs, Rong:2018okz, Atanasov:2018kqw,Atanasov:2022bpi,Zhou:2025kng}, which may be viewed as a supersymmetric version of the $3d$ Ising CFT.  

In \cite{Giombi:2021cnr}, the boundary critical behavior of the Gross-Neveu and Gross-Neveu-Yukawa  models was investigated using both large $N$ methods and the $4-\epsilon$ expansion. By formulating the problem on hyperbolic space, three conformal boundary phases were identified. Ordered from most to least stable under boundary deformations, these conformal boundary phases were denoted in \cite{Giombi:2021cnr} as $B_1$, $B_2$ and $B_2'$, and they can be viewed respectively as the analogs of the normal (or extraordinary), ordinary, and special boundary conditions. Various observables, including some boundary scaling dimensions and free energies, were computed at leading order at large $N$ and in the $\epsilon$ expansion. 
 
The purpose of the present paper is to extend this analysis beyond leading order. While the leading-order treatment establishes the existence and qualitative properties of the boundary fixed points, higher-order corrections provide more nontrivial tests and comparisons between different expansion schemes, and should also provide better estimates for the physical observables in the physical dimension and for relevant low values of $N$. With these motivations in mind, we study the first subleading corrections both in the large $N$ expansion of the GN model in general $d$, and in the GNY description near four dimensions. Along the way, we develop efficient methods based on AdS harmonic analysis and boundary conformal block decompositions that may be useful for the study of other boundary conformal field theories.

At large $N$, we compute the first subleading (order $N^0$) contributions to the AdS free energy of the various boundary phases, and also derive the order $1/N$ correction to the dimension of the lowest boundary fermionic operator in the normal phase, extending to fermions the equation of motion method used in \cite{Giombi:2020rmc}. In the $4-\epsilon$ expansion, we evaluate the boundary free energy and scalar one-point function to higher order and compare the resulting expressions with the large $N$ predictions. Correctly accounting for the contribution of curvature counterterms in the $\epsilon$ expansion turns out to be crucial to recover agreement with the large $N$ expansion.  
The agreement between the two approaches provides a very nontrivial check of the proposed boundary phase structure of the Gross-Neveu and Gross-Neveu-Yukawa theories.

The rest of the paper is organized as follows. In section~\ref{normalreview}, we review the three boundary universality classes of the GNY CFT, in both the large-$N$ and $\epsilon$ expansions. In section~\ref{secN}, we compute the AdS free energy for each of the three boundary conditions to order $N^0$ using the spectral method reviewed in Appendix~\ref{app:harmonic}, and determine the boundary anomalous dimension of the leading fermionic operator in the normal universality class by generalizing the equation of motion method \cite{Giombi:2020rmc}. In section~\ref{Sec:Eps_Expansion}, we compute the higher order $\epsilon$ expansion of the bulk one-point $\langle s\rangle$ and the free energy in the normal universality class, and perform the Pad\'e resummation. In particular, we estimate the one-point function coefficient $a_s$ and boundary central charges of the $3d$ super-Ising universality class based on the two-sided Pad\'e approximant.

\section{Review of the boundary universality classes}
\label{normalreview}
In this section, we review the hyperbolic space description of the three conformal boundary conditions of the GNY CFT. \cite{Giombi:2021cnr}
We begin by setting up notations for the $d$ dimensional hyperbolic space EAdS$_d$ and the spinors in EAdS$_d$. In the Poincar\'e coordinates $(x^0, x^i)=(z,\bold{x})$, where $z\ge 0, \bold{x}\in\mathbb R^{d-1}$, the metric of EAdS$_d$ reads 
$ds^2 = \frac{dz^2+d\bold{x}^2}{z^2}$.  Let $\{\gamma^0,\cdots, \gamma^{d-1}\}$ be flat space gamma matrices satisfying $\{\gamma^a, \gamma^b\}= 2\delta^{ab}$. The Dirac operator in EAdS$_d$ can be expressed as $\slashed{\nabla} = z\slashed{\partial}-\frac{d-1}{2}\gamma_0$, where $\slashed{\partial} =  \gamma^a\partial_a$.

Given a free massive fermion $\Psi$ of mass $m$, the propagator $G_\Psi$ satisfies the equation of motion $(\slashed{\nabla}+m)G_\Psi(x_1, x_2) = - \delta^d(x_1, x_2)$.
Imposing the boundary conditions 
\begin{align}\label{fbc}
    \gamma^0\Psi(z\to 0,\bold{x})=\mp \,{\rm sign}(m)\,\Psi(z\to 0,\bold{x})~,
\end{align}
the boundary spectrum of $G_\Psi$ contains a single fermionic operator of dimension $\hat\Delta_\pm = \frac{d-1}{2}\pm |m|$. The ``$-$'' sign in \eqref{fbc} corresponds to the standard quantization and the ``+'' sign corresponds to the alternative quantization. The explicit fermion propagator with the $\hat\Delta_{\pm}$ boundary condition is \cite{Mueck:1999efk, Basu:2006ti, Giombi:2021cnr}
\begin{equation}\label{Gm}
 \begin{split}
    G_{m}^{\pm }(x_1,x_2) &= \frac{-\left(\frac{1}{2}\pm |m|\right)_{\frac{d-1}{2}}}
{4^{1\pm |m|}(4\pi)^{\frac{d-1}{2}}}
\bigg[
\frac{\slashed{x}_{12}}{\sqrt{z_1 z_2}}
\frac{_2 F_1\left(1\pm |m|-\tfrac{d}{2},\,1\pm |m|,\,1\pm 2|m|,\,-\tfrac{1}{\xi}\right)}{\xi^{1\pm |m|}(1+\xi)^{\frac{d}{2}-1}}\\
&\mp \,\mathrm{sgn}(m)\,
\frac{\gamma_0\slashed{\bar{x}}_{12}}{\sqrt{z_1 z_2}}
\frac{_2 F_1\left(1\pm |m|-\tfrac{d}{2},\,\pm |m|,\,1\pm 2|m|,\,-\tfrac{1}{\xi}\right)}{\xi^{\pm |m|}(1+\xi)^{\frac{d}{2}}}
\bigg]~,
 \end{split}   
\end{equation}
where $x_{12} = (z_{12}, \bold{x}_{12})$, $\bar x_{12} = (-z_1-z_2, \bold{x}_{12})$ and the $\SO(1,d)$ invariant cross-ratio
\begin{gather}\label{xidef}
   \xi=\frac{z_{12}^2+\bold{x}_{12}^2}{4z_1 z_2}~.
 \end{gather}
 
For completeness, let us also recall the free energy of a free massive
fermion on EAdS$_d$. It is given by the following spectral integral
\cite{Giombi:2021cnr}
\begin{equation}
   F_{\Psi}
   =
   -\tr \log\left(\slashed{\nabla}+m\right)
   =
   -\frac{c_d{\rm V}_d}
    {(4\pi)^{d/2}\Gamma\left(\frac d2\right)}
    \int_0^\infty d\lambda\,
    \left|
    \frac{\Gamma\left(\frac d2+i\lambda\right)}
         {\Gamma\left(\frac12+i\lambda\right)}
    \right|^2
    \log\left(\lambda^2+m^2\right) ,
\end{equation}
where ${\rm V}_d$ is the regularized volume of EAdS$_d$. As explained in
\cite{Giombi:2021cnr}, this expression can be evaluated in arbitrary
dimension $d$ by differentiating with respect to $m$, performing the resulting spectral integral by summing residues, and then integrating back in $m$. This
gives
\begin{equation}
\label{FreeEnergyMassiveFermionEqSec2}
    F_{\Psi}
    =
    F_{\rm free}
    -
    c_d\frac{\Gamma \left(1-\frac{d}{2}\right){\rm V}_d}
    {(4 \pi )^{d/2}}
    \int_0^m d\mu\,
    \frac{\Gamma \left(\frac{d}{2}+\mu \right)}
    {\Gamma \left(1-\frac{d}{2}+\mu \right)} ,
\end{equation}
where $F_{\rm free}$ is the free energy of a massless free
fermion, reviewed in Appendix~\ref{FreeEnergyMasslessFermions}.

\subsection{Large-$N$ expansion}
Placing the GN action \eqref{GNaction} in EAdS$_d$ and integrating out the fermions yields the free energy 
\begin{align}
F_{\rm GN} = -\log\int D\sigma e^{N\tr\log (\slashed{\nabla}+\sigma)}.
\end{align}
In the large $N$ limit, evaluating the path integral over $\sigma$ is equivalent to solving the saddle point. Assuming a constant configuration of $\sigma$ \footnote{In the large $N$ limit, the $\sigma$ operator of the GN CFT has dimension 1. Placing the GN CFT in the half space $\mathbb R^{d-1}\times \mathbb R_{\ge 0}$, the one-point function of $\sigma$ is proportional to $1/z$ when the boundary is conformal, where $z$ denotes the distance to the boundary. As we map the half space to the hyperbolic space by a Weyl transformation, the $z$ dependence in $\langle\sigma\rangle$  drops out accordingly. }, the saddle point equation reads $\tr\left(\frac{1}{\slashed{\nabla}+\sigma_\star}\right)  =0$,
where ``$\tr$'' depends on the boundary condition of the fermions.
The normal fixed point corresponds to choosing the standard boundary condition for all $\Psi_I$. In this case, the saddle point equation becomes
\begin{align}\label{saddlenormal}
 \frac{\operatorname{sgn}(\sigma^*)c_d {\rm V}_d}{(4\pi)^{\frac{d}{2}}} \frac{\Gamma\left(1 - \frac{d}{2}\right)\Gamma\left(\frac{d}{2} + |\sigma^*|\right)}{\Gamma\left(1 - \frac{d}{2} + |\sigma^*|\right)}=0~,
\end{align}
and admits the unique solution $|\sigma_\star|  =\frac{d}{2}-1$ provided $2<d<4$. The scaling dimension of the boundary fermion is thus $\hat\Delta_{(1/2)} = d-\frac{3}{2}$. In section \ref{sec:anomdim}, we will compute the $1/N$ correction of $\hat\Delta_{(1/2)}$.
The different signs of $\sigma_\star$ are related by parity. In this paper, without loss of generality, we take $\sigma_\star =\frac{d}{2}-1 $. The fermion propagator becomes an elementary function at this special mass
\begin{align}\label{GPsi^nor}
    G_\Psi^{\rm nor}(x_1,x_2) = 
-\frac{\Gamma\!\left(\frac{d}{2}\right)}
{2\,(4\pi)^{\frac{d}{2}}
\sqrt{z_1 z_2}}
\frac{
\slashed{x}_{12}(1+\xi) 
- \gamma_0\,
\slashed{\bar{x}}_{12}\,\xi}{\left(\xi(1+\xi)\right)^{\frac{d}{2}}}~.
\end{align}
It corresponds to setting $m=\frac{d}{2}-1$ and choosing the ``+'' sign in \eqref{Gm}.

Consider the fluctuations of $\sigma$ around the saddle point $\sigma=\sigma_\star+ \delta\sigma$. The quadratic action of $\delta\sigma$ is induced by a fermion loop \begin{tikzpicture}[baseline=-0.1cm]
    \draw [dashed](-0.8,0) -- (-0.3,0);
    \draw [dashed](0.3,0) -- (0.8,0);
    \draw (0,0) circle (0.3cm);
\end{tikzpicture}:
\begin{equation}\label{S2sigma}
    \begin{split}
         &S_{2,\sigma} =\frac{1}{2} \int d^d x_1 d^d x_2\sqrt{g_{x_1}} \sqrt{g_{x_2}}\delta\sigma(x_1) \CK_\sigma(x_1,x_2) \delta\sigma(x_2)\,,\\
         & \CK_\sigma(x_1,x_2) = -\frac{N\,\Gamma\!\left(\frac{d}{2}\right)^2 c_d}{(4\pi)^d }  \frac{1}{\left(\xi(1+\xi)\right)^{d-1}}\,~.
    \end{split}
\end{equation}
Inverting $\CK_\sigma$ yields the two-point function of $\delta\sigma$
 \begin{align}\label{Gsigmanorm}
     G^{\rm nor}_{\sigma}(\xi) = -\frac{2^{2d-5}(d-2)\,\Gamma\!\left(\frac{d-1}{2}\right)^2\Gamma(d)}
{N c_d\,\pi\,\Gamma\!\left(\frac{d}{2}\right)
\Gamma\!\left(1-\frac{d}{2}\right)
\Gamma(2d-2)\,\xi^d}
\;{}_2F_1\!\left(d,\,d-1,\,2d-2,\,-\frac{1}{\xi}\right) 
 \end{align}
 from which we can read off the boundary spectrum of  $\delta\sigma$. It consists of boundary operators with dimension $d+2n, n\ge 0$. The leading operator is  the displacement operator. There is thus no O$(N)$ invariant relevant boundary deformation at the normal fixed point.  

With the alternative boundary condition imposed on the fermions, the saddle point equation is
\begin{align}\label{saddleB2}
 \frac{\operatorname{sgn}(\sigma^*)c_d {\rm V}_d}{(4\pi)^{\frac{d}{2}}} \frac{\Gamma\left(1 - \frac{d}{2}\right)\Gamma\left(\frac{d}{2} - |\sigma^*|\right)}{\Gamma\left(1 - \frac{d}{2} - |\sigma^*|\right)}=0~.
\end{align}
It 
admits a unitary  saddle $|\sigma_\star| = 2-\frac{d}{2}$ for $3\le d<4$. This saddle point corresponds to both ordinary and special phases. At the leading order of large $N$ expansion, the fermion propagator takes the same form in these two phases 
\begin{align}\label{GPsi^ord}
    G_\Psi^{\rm ord}(x_1,x_2) = 
 -\frac{\Gamma\!\left(\frac{d}{2}-1\right)}{4\pi^{\frac{d}{2}}\sqrt{z_1 z_2}}
\frac{\slashed{x}_{12}\left((\xi+1)(d-2+2(d-3)\xi)\right)
+\gamma_0\slashed{\bar x}_{12}\,\xi\left(d-4+2(d-3)\xi\right)}{\left(\xi(1+\xi)\right)^{\frac{d}{2}}}~,
\end{align}
where we have chosen $\sigma_\star = 2- \frac{d}{2}$. The ordinary and special phases  are distinguished by the $\sigma$ propagator, or more precisely, by the leading block in its boundary conformal block expansion. We will not give the full expression of $\sigma$ propagator here; it can be found in \cite{Giombi:2021cnr}. We will return to a more detailed discussion of the leading block in the next section using the method of AdS harmonic analysis.

Before moving to the $\epsilon$-expansion, let us comment on the large $N$ limit in $3d$. As $d\to 3$, the two saddle points $\frac{d}{2}-1$ and $2-\frac{d}{2}$ coincide, and one can  further verify that $G_\Psi^{\rm nor}$ (cf. \eqref{GPsi^nor}) and $G_\Psi^{\rm ord}$ (cf. \eqref{GPsi^ord}) also coincide. Indeed, $\sigma_\star= \frac{1}{2}$ corresponds to the unitarity bound of spinors in 3$d$. At this value, the other solution of the Dirac equation develops a logarithm in $\xi$. It suggests that for sufficiently high $N$, the normal boundary condition is the unique conformal boundary condition in $3d$. For further related discussions, see \cite{Diatlyk:2026gab}.

\subsection{$\epsilon$-expansion}
On EAdS$_d$, the conformal coupling of the scalar to the background
curvature gives rise to the curvature induced mass term
$-\frac{d(d-2)}{8}s_0^2$. In addition, renormalization of the
interacting theory on a curved background requires curvature
counterterms. We therefore write the bare action as
\begin{gather}\label{GNYAdS}
    S=\int d^d x\sqrt{g}\left[
    \frac{1}{2}\left(\partial_\mu s_0\right)^2
    -\frac{d(d-2)}{8}s_0^2
    - \left( \bar{\Psi}^I_0 \slashed{\nabla} \Psi_{0,I}
    + g_{1,0} s_0 \bar{\Psi}_0^I \Psi_{0,I}\right)
    + \frac{g_{2,0}}{4!} s_0^4
    \right]+S_{\rm curv.}~.
\end{gather}
Here, quantities with a subscript ``0'' are bare. The term
$S_{\rm curv.}$ denotes additional curvature counterterms, beyond the
canonical conformal coupling already included in the scalar mass term.
Its explicit form and its contribution to the free energy are discussed
in Section~\ref{SubSec:CurvatureTerm}.

The potential $V(s_0) = -\frac{d(d-2)}{8}s_0^2+\frac{g_{2,0}}{4!}s_0^4$ has a local maximal at $s_0=0$. The scalar quadratic fluctuation around this saddle point describes a conformally coupled scalar in EAdS$_d$, for which we can choose either the Dirichlet or the Neumann boundary condition.
The ordinary phase corresponds to perturbing the Dirichlet boundary condition and the special phase corresponds to perturbing the Neumann boundary condition. The potential $V(s_0)$ has an additional 
minimum at 
\begin{gather}
s_{\star,0}=\sqrt{\frac{3d(d-2)}{2g_{2,0}}}\,.
    \label{s_star}
\end{gather}
The normal boundary universality class corresponds to perturbing this  nontrivial saddle point. Expanding the action \eqref{GNYAdS} around $ s_{\star,0}$ yields 
\begin{gather}
    S=-\frac{d^2(d-2)^2}{32g_{2,0}}\int d^dx\sqrt{g}
    +\int d^dx \sqrt{g}\left(
    \frac{\left(\partial_\mu t\right)^2}{2}
    +\frac{d(d-2)}{4}t^2
    -\bar{\Psi}^I_0 \left(\slashed{\nabla} +\mu_0\right)\Psi_{0,I}\right)\nonumber\\
    +\int d^dx \sqrt{g}\left(- g_{1,0} \,t\, \bar{\Psi}^I_0 \Psi_{0,I}
    + \sqrt{\frac{3d(d-2)g_{2,0}}{2}} \, \frac{ t^3}{6}
    +\frac{g_{2,0}}{4!} t^4
    \right)\,,
\end{gather}
where $t$ is the fluctuation of $s$, i.e.   $s_0=s_{\star,0}+t$, and $\mu_0$ is the bare fermion mass induced  by the nonzero saddle point of $s$
\begin{gather}
    \mu_0=g_{1,0}s_{\star,0}
    =\frac{g_{1,0}}{\sqrt{g_{2,0}}}\sqrt{\frac{3d(d-2)}{2}}\,.
    \label{BareFermionMass}
\end{gather}
Close to $d=4$ dimensions, for both $t$ and $\Psi^I$, only the standard quantization is unitary. So, the free fermion propagator is $G^+_{\mu_0}$, c.f. \eqref{Gm}, and the free scalar propagator is 
\begin{align}
    G_t(x_1,x_2)
    =\frac{\Gamma\left(\hat{\Delta}_t\right)}{2^{\hat{\Delta}_t+1}\pi^{\frac{d-1}{2}}\Gamma\left(\hat{\Delta}_t+\frac{3-d}{2}\right)}
    \frac{1}{u^{\hat{\Delta}_t}}
    {}_2 F_1\left(\frac{\hat{\Delta}_t}{2},\frac{\hat{\Delta}_t+1}{2},\hat{\Delta}_t+\frac{3-d}{2},\frac{1}{u^2}\right),
    \label{scalar_t_GreenFunction}
\end{align}
where $u\equiv 2\xi+1$ and $\hat{\Delta}_t=\frac{d-1+ \sqrt{3 d^2-6 d+1}}{2}$. For later convenience, we introduce the following short-hand notations of two-point functions in the coincident limit. For the fermions, summing up both flavor and spinor indices in the coincident limit yields 
\begin{gather}
    \langle \bar{\Psi}_0 \Psi_{0}\rangle
    \equiv Nc_d\frac{  \Gamma \left(1-\frac{d}{2}\right) \Gamma \left(\frac{d}{2}+\mu_0 \right)}{(4 \pi )^{d/2} \Gamma \left(1-\frac{d}{2}+\mu_0 \right)}\,.
    \label{Gpsi(x,x)}
\end{gather}
Similarly, for the scalar field $t$, we have
\begin{gather}
    G_t(1)=\frac{\Gamma\left(1-\frac{d}{2}\right)\Gamma\left(\hat{\Delta}_t\right)}{(4\pi)^{\frac{d}{2}}\Gamma\left(2-d+\hat{\Delta}_t\right)}\,.
    \label{Gt(x,x)}
\end{gather}

\section{The leading $1/N$ corrections}\label{secN}
In the previous section, we reviewed the realization of the three boundary phases in the leading order of large-$N$ expansion. In this section, we compute the $1/N$ correction to the free energy in these boundary phases, and the boundary fermion anomalous dimension in the normal phase. We also carry out similar free energy calculations for various conformal boundary conditions of the Wilson-Fisher CFT.

\subsection{Free energy of the GNY model at order $N^0$}\label{Free}
\subsubsection{The normal phase }
The fermion one-loop path integral contributes to the free energy at order $N$. The subleading contribution, denoted $F^{(1)}_{\rm nor}$, comes from the one-loop free energy of $\delta\sigma$. Integrating out $\delta\sigma$ in \eqref{S2sigma} leads to $ F^{(1)}_{\rm nor} = \frac{1}{2}\log\det\CK_\sigma$.
 As reviewed in Appendix~\ref{app:harmonic}, the functional determinant of any AdS-invariant two-point function $G$ is entirely encoded in the spectral density $\rho_G(\nu)$, c.f. \eqref{logdet_general} \footnote{We use the shorthand notation $\Gamma(a\pm b) \equiv \Gamma(a+b)\Gamma(a-b)$.}:
\begin{align}\label{logdet_general1}
\frac{1}{2}\log\det(G) = \frac{{\rm V}_d}{(4\pi)^{d/2}\Gamma(\frac{d}{2})}\int_0^\infty d\nu\, \frac{\Gamma(\frac{d-1}{2}\pm i\nu)}{\Gamma(\pm i\nu)}\log\rho_G(\nu)~,
\end{align}
where $\frac{\Gamma(\frac{d-1}{2}\pm i\nu)}{\Gamma(\pm i\nu)}$ is the Plancherel measure of the $\SO(1,d )$ group and ${\rm V}_d$ denotes the regularized volume of the $d$-dimensional Euclidean AdS  of unit radius.
Therefore, computing $F^{(1)}_{\rm nor}$ boils down to two  steps: (i) calculating the spectral density of $\CK_\sigma$ using \eqref{rhoG}, and (ii) evaluating \eqref{logdet_general1} with $\rho_{\CK_\sigma}$ obtained in  step (i).

For step (i), we first express $\CK_\sigma$ in terms of $r = {\rm arccosh}(2\xi+1)$, 
\begin{align}\label{CGdefGNY}
\CK_\sigma(x_1,x_2) = -\frac{Nc_d\,\Gamma\!\left(\frac{d}{2}\right)^2 }{4\pi^d \sinh(r)^{2d-2}}~. 
\end{align}
The corresponding spectral density follows from \eqref{rhoG}, and \eqref{Sk_result} with 
 $\kappa = 2-2d+ (d-1) = 1-d$:
 \begin{align}\label{rho_ord}
 \rho_{\CK_\sigma}(\nu) = -\frac{N c_d\pi^{\frac{3}{2}}\,  \hat g_{\frac{3(d-1)}{4},\frac{d+1}{4}}(\nu)}{(4\pi)^{\frac{d}{2}}\sin(\frac{d\pi}{2})\Gamma(\frac{d-1}{2})}~, \qquad \hat g_{a, b}(\nu)\equiv\frac{\Gamma(a\pm i\frac{\nu}{2})}{\Gamma(b\pm i\frac{\nu}{2})}~.
 \end{align}
It is positive for $2<d<4$.

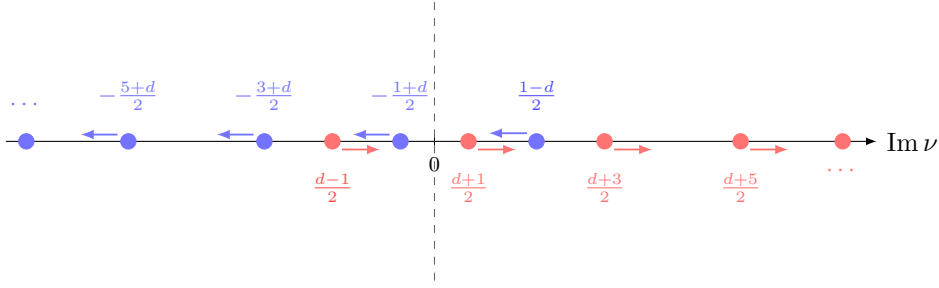
\begin{figure}[t]
    \centering
\begin{tikzpicture}[scale=0.90, >=latex]
 
  \draw[->] (-6.3,0) -- (6.5,0)
      node[right, font=\small] {$\operatorname{Im}\nu$};
 
  \draw (0,0.06) -- (0,-0.06);
  \node[below=2pt, font=\scriptsize] at (0,0) {$0$};
 
  \draw[dashed, gray!120] (0,-2.05) -- (0,2.05);
 
 
  \filldraw[blue!55] ( 1.5, 0) circle (3.2pt);
  \draw[->, blue!55, line width=0.7pt] (1.36,0.12) -- (0.8,0.12);
  \node[above=5pt, blue!70, font=\footnotesize] at (1.5,0.12)
      {$\frac{1-d}{2}$};
 
  \filldraw[blue!55] (-0.5, 0) circle (3.2pt);
  \draw[->, blue!55, line width=0.7pt] (-0.66,0.10) -- (-1.2,0.10);
  \node[above=5pt, blue!55, font=\footnotesize] at (-0.5,0.12)
      {$-\frac{1+d}{2}$};
 
  \filldraw[blue!55] (-2.5, 0) circle (3.2pt);
  \draw[->, blue!55, line width=0.7pt] (-2.66,0.10) -- (-3.20,0.10);
  \node[above=5pt, blue!55, font=\footnotesize] at (-2.5,0.12)
      {$-\frac{3+d}{2}$};
 
  \filldraw[blue!55] (-4.5, 0) circle (3.2pt);
  \draw[->, blue!55, line width=0.7pt] (-4.66,0.10) -- (-5.20,0.10);
  \node[above=5pt, blue!55, font=\footnotesize] at (-4.5,0.12)
      {$-\frac{5+d}{2}$};
 
  \filldraw[blue!55] (-6.0, 0) circle (3.2pt);
  \node[above=5pt, blue!55, font=\footnotesize] at (-6.0,0.12)
      {$\cdots$};
 
 
  \filldraw[red!55] (-1.5, 0) circle (3.2pt);
  \draw[->, red!55, line width=0.7pt] (-1.36,-0.12) -- (-0.8,-0.12);
  \node[below=5pt, red!70, font=\footnotesize] at (-1.5,-0.12)
      {$\frac{d-1}{2}$};
 
  \filldraw[red!55] ( 0.5, 0) circle (3.2pt);
  \draw[->, red!55, line width=0.7pt] (0.64,-0.12) -- (1.2,-0.12);
  \node[below=5pt, red!55, font=\footnotesize] at (0.5,-0.12)
      {$\frac{d+1}{2}$};
 
  \filldraw[red!55] ( 2.5, 0) circle (3.2pt);
  \draw[->, red!55, line width=0.7pt] (2.64,-0.12) -- (3.2,-0.12);
  \node[below=5pt, red!55, font=\footnotesize] at (2.5,-0.12)
      {$\frac{d+3}{2}$};
 
  \filldraw[red!55] ( 4.5, 0) circle (3.2pt);
  \draw[->, red!55, line width=0.7pt] (4.64,-0.12) -- (5.20,-0.12);
  \node[below=5pt, red!55, font=\footnotesize] at (4.5,-0.12)
      {$\frac{d+5}{2}$};
 
  \filldraw[red!55] ( 6.0, 0) circle (3.2pt);
  \node[below=5pt, red!55, font=\footnotesize] at (6.0,0)
      {$\cdots$};
 
\end{tikzpicture}
\caption{Pole structure of the Plancherel measure when $-1<d<0$. The blue dots correspond to the poles of $\Gamma(\frac{d-1}{2}+i\nu)$ and the red dots correspond to the poles of $\Gamma(\frac{d-1}{2}-i\nu)$. The arrows show how the poles move as we increase $d$. When $d>1$, the leftmost red dot and the rightmost blue dot cross the contour (the dashed line).  }
\label{polecrossing}
\end{figure}

We then carry out step (ii). Substituting $\rho_{\CK_\sigma}$ into \eqref{logdet_general1} gives
\begin{align}\label{Fsigma}
F_{\rm nor}^{(1)} = -\frac{{\rm V}_d}{(4\pi)^{d/2}\Gamma(\frac{d}{2})}\int_0^\infty d\nu\, \frac{\Gamma(\frac{d-1}{2}\pm i\nu)}{\Gamma(\pm i\nu)} \log \hat g_{\frac{d+1}{4}, \frac{3d-3}{4}}(\nu)~.
\end{align}
The $\nu$-independent prefactor in  $\rho_{\CK_\sigma}$ does not contribute to the free energy due to  $\int \limits_{0}^{\infty} d\nu\, \frac{\Gamma(\frac{d-1}{2}\pm i\nu)}{\Gamma(\pm i\nu)} = 0$ in dimensional regularization. Because we will encounter similar integrals, it is useful to explain this fact in more detail.
Consider $-1<d<0$, so  the integral of the Plancherel measure  is finite. Since the integrand is an even function in $\nu$, we extend the integral domain to the whole real line. We evaluate the resulting integral by closing the contour in the upper half plane. The corresponding poles are thus $\nu = i\left(\frac{d-1}{2}+n\right), n\ge 1$ and $\nu = - i \frac{d-1}{2}$. Summing up the residues yields 
\begin{align}\label{muint1}
-1<d<0: \quad \int_{-\infty}^\infty d\nu\, \frac{\Gamma(\frac{d-1}{2}\pm i\nu)}{\Gamma(\pm i\nu)} = -2\cos\left(\frac{\pi d}{2}\right)\Gamma(d)~.
\end{align}
Then we analytically continue this result to the physically interesting dimensions, i.e. $d\ge 2$. Note that as $d$ crosses 1, the pole at $\nu = i \frac{d-1}{2}$ enters the upper half plane while the pole at $\nu = -i \frac{d-1}{2}$ exits the upper half plane. We illustrate the pole crossing pattern in Figure \ref{polecrossing}.  By adding  the residue at $i \frac{d-1}{2}$ and subtracting the residue at $-i \frac{d-1}{2}$, we get 
\begin{align}\label{muint2}
d>1: \quad \int_0^\infty d\nu\, \frac{\Gamma(\frac{d-1}{2}\pm i\nu)}{\Gamma(\pm i\nu)} = 0~.
\end{align}

For practical purposes, such as  computing the $\log(R)$ coefficient in $d=3$, or extracting the $\epsilon$-expansion around 4 dimensions, \eqref{Fsigma} is not very convenient. An obvious reason is that expanding the integrand in $\epsilon$ first does not lead to a convergent integral term by term. To overcome such technical difficulties, we introduce an auxiliary function
\begin{align}\label{Fx}
\Phi(x) \equiv  \int_{\mathbb R} d\nu\, \frac{\Gamma(\frac{d-1}{2}\pm i\nu)}{\Gamma(\pm i\nu)} \log \Gamma\!\left(x\pm i\frac{\nu}{2}\right)~.
\end{align}
By construction, $F_{\rm nor}^{(1)}$ is proportional to the integral of $\Phi'(x)$ from $\frac{3d-3}{4}$ to $\frac{d+1}{4}$.
There are two advantages of using $\Phi'(x)$. First, it admits a simple analytical expression. Second, the non-compact integral domain of $\nu$ becomes a compact integral domain of $x$. To show the first point, we compute $\Phi'(x)$ as follows:
\begin{align}\label{dFx}
\Phi'(x)= 2\int_{\mathbb R} d\nu\, \frac{\Gamma(\frac{d-1}{2}\pm i\nu)}{\Gamma(\pm i\nu)} \psi\!\left(x- i\frac{\nu}{2}\right)~,
\end{align}
where $\psi(z)=\Gamma'(z)/\Gamma(z)$. Using the integral representation of $\psi(z)$
\begin{align}\label{psiz}
\psi(z) = \int_0^\infty dt\, \left(\frac{e^{-t}}{t}-\frac{e^{- z t}}{1-e^{-t}}\right)~, \qquad \Re(z)>0~,
\end{align}
and exchanging the order of integration gives
\begin{align}\label{dF1}
\Phi'(x)=2 \int_0^\infty dt\, \left[\frac{e^{-t}}{t}\int_{\mathbb R} d\nu\, \frac{\Gamma(\frac{d-1}{2}\pm i\nu)}{\Gamma(\pm i\nu)}-\frac{e^{- x t}}{1-e^{-t}} \int_{\mathbb R} d\nu\, \frac{\Gamma(\frac{d-1}{2}\pm i\nu)}{\Gamma(\pm i\nu)}e^{i\frac{\nu}{2} t}\right]~.
\end{align}
We again consider $-1<d<0$, so both $\nu$ integrals in \eqref{dF1} are convergent. The first integral has been discussed above and its value is  given by \eqref{muint1}. The second integral can be evaluated similarly by closing the contour in the upper half plane
\begin{align}\label{FPlan}
 \int_{\mathbb R} d\nu\, \frac{\Gamma(\frac{d-1}{2}\pm i\nu)}{\Gamma(\pm i\nu)}e^{i\frac{\nu}{2} t} =\cos\!\left(\frac{d\pi}{2}\right)\Gamma(d)\left[\frac{(1+e^{-\frac{t}{2}})e^{-\frac{d-1}{4}t}}{(1-e^{-\frac{t}{2}})^d}-\left(e^{\frac{d-1}{4}t}+e^{-\frac{d-1}{4}t}\right)\right]~.
\end{align}
Combining \eqref{muint1} and \eqref{FPlan}, and using the digamma representation \eqref{psiz}, we obtain 
\small
\begin{align}\label{Fprime_neg}
-1<d<0:\quad \Phi'(x) = -2\cos\!\left(\frac{d\pi}{2}\right)\Gamma(d) \left[ \psi\!\left(x\!+\!\frac{d\!-\!1}{4}\right)\!+\!\psi\!\left(x\!-\!\frac{d\!-\!1}{4}\right)\right]+\frac{2\pi}{d\sin(\frac{d\pi}{2})}\frac{\Gamma(2x-\frac{1-d}{2})}{\Gamma(2x-\frac{1+d}{2})}~.
\end{align}
\normalsize
The analytical continuation to $d\ge 2$ is exactly the same as the discussions that lead to \eqref{muint2} from \eqref{muint1}. There is a similar pole crossing pattern for both $\nu$ integrals in \eqref{dF1}. Taking the pole crossing  into account amounts to discarding the digamma functions in \eqref{Fprime_neg}. Altogether, we have 
\begin{align}\label{Fprime_final}
d\ge 2: \quad \Phi'(x) =\frac{2\pi}{d\sin(\frac{d\pi}{2})}\frac{\Gamma(2x-\frac{1-d}{2})}{\Gamma(2x-\frac{1+d}{2})}~.
\end{align}
Plugging \eqref{Fprime_final} into \eqref{Fsigma} leads to 
\begin{align}\label{Fsigmaord}
F_{\rm nor}^{(1)} = \frac{\Gamma(-\frac{d}{2}){\rm V}_d}{2(4\pi)^{\frac{d}{2}}}\int_{\frac{3d-3}{4}}^{\frac{d+1}{4}} dx\, \frac{\Gamma \left(2x+\frac{d-1}{2} \right)}{\Gamma \left(2x-\frac{d+1}{2}\right)}=\Gamma (-d)\int_{\frac{3d-3}{4}}^{\frac{d+1}{4}} dx\, \frac{\Gamma \left(2x+\frac{d-1}{2} \right)}{\Gamma \left(2x-\frac{d+1}{2}\right)}~,
\end{align}
where we have used the the dimensional regularized volume ${\rm V}_d =\pi^{\frac{d-1}{2}}\Gamma(\frac{1-d}{2})$.
This representation is particularly suitable for extracting the $\epsilon$-expansion around 4 dimensions. Substituting $x = (1-\frac{d}{2})u+ \frac{3d-3}{4}$, the domain of the integral is mapped to the interval $0\le u\le 1$. We can expand the integrand in $\epsilon$ and perform the $u$ integral for each term. Up to order $\epsilon$, the $\epsilon$-expansion reads
\begin{equation}\label{FB1order1}
\begin{split}
F_{\rm nor}^{(1)}\approx &-\frac{269}{180 \epsilon }+\frac{1}{72} \left(222 \log (A)-12 \zeta '(-3)-\frac{45 \zeta
   (3)}{2 \pi ^2}+\frac{9691}{40}+\frac{538 \gamma_E }{5}-90 \log (2 \pi )\right)\\
  & -4.55964
   \epsilon +O\left(\epsilon ^2\right)~.
   \end{split}
\end{equation}

In $3d$, the regularized volume of EAdS$_3$ is given by ${\rm V}_3 = -2\pi\log (R)$, where $R$ is the radius of the $S^2$ at a large IR cutoff. Setting $d=3$ in \eqref{Fsigmaord} gives
\begin{align}
    3d: \quad F_{\rm nor}^{(1)} = \log(R)\int_1^{\frac{3}{2}}\frac{2}{3}x(x-1)(2x-1) = \frac{3}{16}\log(R)~.
\end{align}
On the other hand, the one-loop free energy of a single two-component Dirac fermion with boundary dimension $\hat\Delta = \frac{3}{2}$ is $\frac{\log(R)}{12}$ \cite{Giombi:2021cnr}.
Therefore, the boundary central charge of the normal phase in the large $N$ expansion is 
\begin{align}
    c_{\rm nor} = - \frac{N}{4}-\frac{9}{16} + O(1/N)~.
\end{align}
Note that in this result $N$ denotes the number of two-component Dirac fermions, hence here the special case of the 3d super-Ising CFT corresponds to formally setting $N=1/2$.

\subsubsection{The ordinary and special phases}
The quadratic action of $\sigma$ fluctuations in the ordinary and special phases can be schematically written as $S_\sigma \sim \int \delta\sigma(x)\tilde\CK_\sigma(x, y)\delta\sigma$, where $\tilde\CK_\sigma\sim \tr\left( G^{\rm ord}_\Psi(x, y)^2\right)$. Given the explicit expression of $G^{\rm ord}_\Psi$, cf. \eqref{GPsi^ord}, we find the spectrum density of $\tilde\CK_\sigma$: 
\begin{align}
    \rho_{\tilde\CK_\sigma} (\nu) \propto \left(\nu^2+\left(\frac{d-5}{2}\right)^2\right)\hat g_{\frac{3d-7}{4}, \frac{d+1}{4}}(\nu)~,
\end{align}
where the $\nu$-independent overall normalization is suppressed. The spectral density of the $\delta\sigma$ two-point function is thus $\rho_\sigma (\nu)= 1/\rho_{\tilde\CK_\sigma} (\nu)$.
Notice that $\rho_\sigma$ contains $1/(\nu^2+(\frac{d-5}{2})^2)$,
which is the spectral density of a free scalar in EAdS$_d$. Provided $3<d<4$, we can choose either standard or alternate boundary conditions for this factor. The former yields the boundary dimension $\hat\Delta = \frac{d-1}{2}+\frac{5-d}{2} = 2$, corresponding to the ordinary phase,  and the latter yields the boundary dimension $\hat\Delta = \frac{d-1}{2}+\frac{d-5}{2} = d-3$, corresponding to the special phase. The pole structure of the remaining factor $\hat g_{\frac{d+1}{4}, \frac{3d-7}{4}}(\nu)$ indicates boundary operators of dimension $\hat\Delta = d+2n$, with $n\in\mathbb N$ for both phases.
The free energies $F^{(1)}_{\rm ord}$ and $F^{(1)}_{\rm spec}$  are thus related by a double-trace deformation. This double trace structure was checked against $\epsilon$ expansion in \cite{Giombi:2021cnr}. Below, we show that $F^{(1)}_{\rm ord}$ itself is also consistent with the $\epsilon$ expansion derived in \cite{Giombi:2021cnr}.

Substituting $ \rho_{\tilde\CK_\sigma}(\nu)$ into \eqref{logdet_general} naturally leads to two contributions. The first contribution, due to the $\nu^2+(\frac{d-5}{2})^2$ factor, is the same as the one-loop free energy of a $\hat\Delta=2$ scalar. It admits a simple integral representation \cite{Giombi:2020rmc}, from which we can extract the $\epsilon$ expansion around $4d$
\begin{align}\label{FDel2}
    F_{\hat\Delta=2} = F_D + \frac{{\rm V}_d}{2(4\pi)^{\frac{d}{2}}}\int_{\frac{d}{2}}^{2} d\hat{\Delta}\, \frac{(2\hat{\Delta}-d+1)\Gamma(\hat{\Delta})\Gamma\!\left(1-\frac{d}{2}\right)}{\Gamma(2-d+\hat{\Delta})} = F_D-\frac{\epsilon}{32}+O(\epsilon^2)~.
\end{align}
Here, $F_D$ denotes the one-loop free energy of a conformally coupled scalar with the Dirichlet boundary condition. Its $\epsilon$ expansion was carried out in \cite{Giombi:2025pxx} to order $\epsilon^3$ and see \eqref{FDexp1} for the truncation to order $\epsilon$. In this section, the explicit form of $F_D$ is irrelevant. The $\hat g$ factor in $\tilde\CK_\sigma$ gives rise to the second contribution to $F^{(1)}_{\rm ord}$. Following the same strategy outlined in the normal phase case, we can also transform it to an integral with a finite domain:
\begin{align}\label{Ford2}
\frac{\Gamma(-\frac{d}{2}){\rm V}_d}{2(4\pi)^{\frac{d}{2}}}\int_{\frac{3d-7}{4}}^{\frac{d+1}{4}} dx\, \frac{\Gamma \left(2x+\frac{d-1}{2} \right)}{\Gamma \left(2x-\frac{d+1}{2}\right)} = -\frac{\epsilon}{16}+O(\epsilon^2)~. 
\end{align}
Combining \eqref{FDel2} and \eqref{Ford2}, we obtain 
\begin{align}
    F^{(1)}_{\rm ord} = F_D - \frac{3}{32}\epsilon+O(\epsilon^2)~.
\end{align}
The free energy of the ordinary phase for any finite $N$ was computed to order $\epsilon$ in \cite{Giombi:2021cnr}
\begin{align}\label{Fordexp}
    F^{\epsilon\text{-exp}}_{\rm ord} = N F_{\rm free} + F_D+\frac{\epsilon}{6(2N+3)}\left[N^2+\frac{3N}{8}+\frac{\sqrt{4N^2+132N+9}-2N+3}{96}\right]+O(\epsilon^2)~.
\end{align}
Here, $F_{\rm free}$ is the free energy of a massless Dirac fermion. Its $\epsilon$ expansion is given by \eqref{Ffree}. In the large $N$ expansion of $F^{\epsilon\text{-exp}}_{\rm ord}$, the $N^0$ term is exactly $F_D - \frac{3}{32}\epsilon$, up to $O(\epsilon^2)$ corrections.

\subsection{Free energy of the O$(N)$ model at order $N^0$}\label{ONreview}
The method described above is not specific to the normal boundary condition of the GN model. We illustrate its application in computing the free energy of the critical O$(N)$ model with various conformal boundary conditions. We  use $\CF$ to represent free energy of the O$(N)$ model.

The  action of the critical O$(N)$ model in EAdS$_d$ is given by  \cite{Giombi:2020rmc}
\begin{align}
S = \frac{1}{2}\int d^d x \sqrt{g}\left(\frac{1}{2}(\partial\phi^I)^2-\frac{d-2}{8}\phi^I\phi^I +\frac{1}{2}\sigma\phi^I\phi^I \right)~.
\end{align}
It admits two O$(N)$ invariant boundary fixed points, corresponding to the ordinary and special transitions, and one O$(N)$ breaking boundary fixed point, corresponding to the normal (or extraordinary) transition. We review the symmetry breaking phase and discuss the corresponding free energy in Appendix. \ref{sec:extraordinary}. Here, we focus on the two O$(N)$ invariant phases.

In the large $N$ limit, the O$(N)$ invariant fixed points are saddle points of the free energy $\CF^{(0)}(\sigma) = \frac{N}{2}\tr\log(-\nabla^2+\sigma-\frac{d(d-2)}{4} )$. The saddle point condition $\partial_\sigma \CF^{(0)}(\sigma)\big|_{\sigma=\sigma_\star}=0$ gives \cite{Giombi:2020rmc}
\begin{align}
&{\rm Ordinary}: \quad \sigma_\star = \frac{(d-2)(d-4)}{4} ~\Rightarrow~ \hat\Delta_\phi = d-2~,\nonumber\\
&{\rm Special}: \quad \sigma_\star = \frac{(d-4)(d-6)}{4} ~\Rightarrow~ \hat\Delta_\phi = d-3~.
\end{align}
At order $N$, the free energy is $\CF^{(0)}(\sigma_\star)$. The leading $1/N$ correction $  \CF^{(1)} $ arises from the one-loop determinant of the $\sigma$ fluctuations around the saddle point, i.e.  $\sigma = \sigma_\star + i \delta\sigma$. \footnote{We have inserted a factor of $i$ here such that the two-point function of $\delta\sigma$ is positive definite. In section \ref{sec:anomdim}, following the convention of \cite{Giombi:2020rmc}, $G_\sigma$ is defined without $i$. } Expanding  $\CF^{(0)}(\sigma)$ to the second order in $\delta\sigma$ generates the quadratic action
 \begin{align}
 S_2 = \frac{N}{4}\int d^d x d^d y \sqrt{g_x} \sqrt{g_y} \,\delta\sigma(x)G_\phi(x, y)^2 \delta\sigma(y)~,
 \end{align}
 where $G_\phi$ is the propagator of $\phi^I$ at the leading order of large $N$ expansion. Thus, the $\sigma$ propagator $G_\sigma$ is the inverse of $\CG\equiv \frac{N}{2}G_\phi^2$. Integrating out $\delta\sigma$ at the Gaussian level produces:
  \begin{align}
  \CF^{(1)} = \frac{1}{2} \log\det\left(\CG\right)~.
  \end{align}

\noindent{}\textbf{The ordinary transition}.
At the ordinary fixed point, $\CG$ takes the form 
\begin{align}\label{CGdefON}
\CG_{\rm ord}(r) = \frac{N\Gamma(\frac{d}{2}-1)^2}{32\pi^d}\sinh^{4-2d}(r)~.
\end{align}
Using \eqref{Sk_result}, we find the corresponding spectral density $\rho_{\rm ord}(\nu) \propto \hat g_{\frac{3d-7}{4},\frac{d+1}{4}}(\nu)$. It differs the spectral density of the ordinary phase of the GNY CFT by the simple factor $\nu^2+(\frac{d-5}{2})^2$.
So, the one-loop free energy is given by \eqref{Ford2}
\begin{align}\label{Ford1}
\CF_{\rm ord}^{(1)} = \frac{\Gamma(-\frac{d}{2}){\rm V}_d}{2(4\pi)^{\frac{d}{2}}}\int_{\frac{3d-7}{4}}^{\frac{d+1}{4}} dx\, \frac{\Gamma \left(2x+\frac{d-1}{2} \right)}{\Gamma \left(2x-\frac{d+1}{2}\right)}~. 
\end{align}
In $3d$, \eqref{Ford1} gives $ \CF_{\rm ord}^{(1)} =\frac{\log R}{48}$.
This result agrees with \cite{Krishnan_2023}, where the O$(N)$ CFT is placed in a 3$d$ hemisphere of radius $R$, with the ordinary boundary condition imposed on the boundary sphere. In our calculation, $R$ arises as an IR regulator of the boundary sphere of the EAdS$_3$.

In $d=4-\epsilon$ dimensions, the $\epsilon$-expansion of $\CF_{\rm ord}^{(1)}$ reads
\begin{align}\label{Fsigmaordexp}
\CF_{\rm ord}^{(1)}=-\frac{\epsilon }{16}+\frac{35 \epsilon ^2}{576}+\left(\frac{281}{6912}-\frac{5 \pi ^2}{288}\right) \epsilon ^3 +O(\epsilon^4)~.
\end{align}
In \cite{Giombi:2025pxx}, the AdS free energy of the O$(N)$ model with the ordinary boundary condition was computed in $d=4-\epsilon$ dimensions to order $\epsilon^3$ for an arbitrary $N$:
\small
\begin{align}\label{F4}
\CF^{\epsilon\text{-exp}}_{\rm ord} &=N F_D +\frac{ N (N+2)}{96 (N+8)} \epsilon+\frac{5 N (N+2) (N^2+29N+132)}{576 (N+8)^3} \epsilon^2\nonumber\\
&+\frac{N (N+2)\epsilon^3}{6912 (N+8)^5}\bigg(8 \left(7+\pi ^2\right) N^4+\left(2409+184 \pi ^2\right) N^3+6 N^2 \left(-720 \zeta (3)+7309+272 \pi ^2\right) \nonumber\\
&+4 N\left(-13392 \zeta (3)+72265+1792 \pi ^2\right)+32 \left(-4752 \zeta (3)+19983+448 \pi ^2\right)\bigg) +O(\epsilon^4)~.
\end{align}
\normalsize
The order $N^0$ term in the large $N$ expansion of $F^{\epsilon\text{-exp}}_{\rm ord} $
precisely reproduces \eqref{Fsigmaordexp}.

\,

\noindent{}\textbf{The special transition}.
At the special fixed point, $\hat\Delta_\phi = d-3$ and $G_\phi\propto \cosh(r)/\sinh^{d-2}(r)$. The spectral density of $\CG = \frac{N}{2}G_\phi^2$ is more complicated
\begin{align}
\rho_{\CG}(\nu)\propto S_{3-d}(\nu)+S_{5-d}(\nu)\propto  \frac{\nu^2+(\frac{d-5}{2})^2}{\nu^2+(\frac{3d-11}{2})^2}\,\hat g_{\frac{3d-7}{4},\frac{d+1}{4}}(\nu)~,
\end{align}
where $S_\kappa(\nu)$ is defined in appendix~\ref{app:integrals}. 

Noticing that $\hat g_{\frac{3d-7}{4},\frac{d+1}{4}}(\nu)$ is the spectral density at the ordinary fixed point, the free energy at the special fixed point naturally splits into two terms:
 \begin{align}\label{spec1}
 \CF^{(1)}_{\rm spec} &=  \CF^{(1)}_{\rm ord} -\frac{{\rm V}_d}{(4\pi)^{d/2}\Gamma(\frac{d}{2})}\int_0^\infty d\nu\, \frac{\Gamma(\frac{d-1}{2}\pm i\nu)}{\Gamma(\pm i\nu)} \log \left(\frac{\nu^2+(\frac{3d-11}{2})^2}{\nu^2+(\frac{5-d}{2})^2}\right)\nonumber~,
 \end{align}
 where $\CF^{(1)}_{\rm ord}$ is given by \eqref{Ford1}. The remaining term corresponds to the difference of the one-loop free energies of two scalars with boundary dimension $\hat\Delta =2d-6$ and $\hat \Delta=2$, respectively. In $d=4-\epsilon$, we find
 \begin{align}
  \CF^{(1)}_{\rm spec} =  -\frac{\epsilon }{16}-\frac{61\epsilon ^2}{576} +O(\epsilon^3) ~,
 \end{align}
in agreement with the direct $\epsilon$-expansion result of \cite{Giombi:2020rmc}.

\,

\noindent{}\textbf{The extraordinary transition}. We briefly mention the result at the extraordinary fixed point: 
 \begin{align}
\CF^{(1)}_{\rm ext} =  -\frac{{\rm V}_d}{(4\pi)^{d/2}\Gamma(\frac{d}{2})}\int_0^\infty d\nu\, \frac{\Gamma(\frac{d-1}{2}\pm i\nu)}{\Gamma(\pm i\nu)} \log \Big[(\nu^2+\tfrac{(d-1)^2}{4})\,\hat g_{\frac{d+1}{4}, \frac{3(d-1)}{4}}(\nu)\Big]~.
\end{align}
A detailed derivation of this equation is presented in Appendix \ref{sec:extraordinary}. 
Compared with \eqref{Fsigma}, the only difference is an additional factor $\nu^2+\tfrac{(d-1)^2}{4}$ in the spectral density, which corresponds to the free energy of a scalar with boundary dimension $\hat \Delta= d-1$. The $\epsilon$-expansion around $4d$ reads
\begin{align}\label{F1ext}
\CF^{(1)}_{\rm ext}=-\frac{4}{3\epsilon}+2.66403-4.29451\epsilon+O\left(\epsilon^2\right)~.
\end{align}
On the other hand, the $\epsilon$-expansion of the free energy of the O$(N)$ model in the extraordinary phase was computed to order $\epsilon$ in \cite{Giombi:2025pxx}
\begin{equation}\label{}
    \begin{split}
    \CF^{\,\epsilon\text{-exp}}_{\rm ext} &= -\frac{30(N+8)-N}{180\epsilon} - \frac{0.0057916N^2 - 2.6177N - 16.312}{N+8}\\
&- \frac{\left(0.26513N^4 + 10.658N^3 + 162.62N^2 + 1097.4N + 2591.9\right)\epsilon}{(N+8)^3} + O(\epsilon^2)~.
\end{split}
\end{equation}
The large $N$ expansion  of $\CF^{\,\epsilon\text{-exp}}_{\rm ext}$ at order $N^0$ is in agreement with \eqref{F1ext}. 

\subsection{Boundary anomalous dimension at order $1/N$}\label{sec:anomdim}

Next, we compute the leading $1/N$ correction to the boundary anomalous dimension of the lowest boundary fermionic operator at the normal boundary fixed point of the GNY CFT. The result 
takes the following remarkably  compact form
\begin{gather}
\hat{\Delta}_{(1/2)}=d-\frac{3}{2}+\frac{1}{N}\,\frac{2^{2d-3}\,\Gamma(d-\frac{1}{2})}{\sqrt{\pi}\,c_d\, d\,\Gamma(d-2)}+O\left(\frac{1}{N^2}\right)~.
\end{gather}
Our analysis adapts the equation-of-motion method developed for the scalar O$(N)$ model in \cite{Giombi:2020rmc} to the present fermionic setting.

\subsubsection{The boundary anomalous dimension in the O$(N)$ CFT}

Let us first briefly recall the logic of this method in the O$(N)$ model \cite{Giombi:2020rmc}. Acting twice with the equation-of-motion operator 
\begin{align}
    \CD=-\nabla^2-\frac{d(d-2)}{4}+\sigma_\star\,, \quad \sigma_\star= \frac{(d-2)(d-4)}{4}\,,
\end{align}
 on $\langle \phi^I(x)\phi^J(y)\rangle$ produces, at order $1/N$, a quartic differential equation of the form 
$\CD^2 G_\phi = G_\phi G_\sigma$.
The two-point function $G_\sigma$ starts at order $1/N$, and its explicit expression was computed in \cite{Giombi:2020rmc}. In terms of the variable $v \equiv\frac{1}{\xi+1}$, it takes the form
\begin{align}
    G_{\sigma}(v)=\frac{1}{N}\frac{(d-4) (d-2) \sin \left(\frac{\pi  d}{2}\right) \Gamma \left(\frac{d-1}{2}\right) \Gamma \left(\frac{d+1}{2}\right)}{\pi ^{3/2}  \Gamma \left(d-\frac{3}{2}\right)}v ^{d}\, _2F_1\left(d,d-2;2 d-4;v\right)\,.
\end{align}
We then expand the  bulk scalar two-point function, including its $1/N$ corrections, in boundary conformal blocks:
\begin{gather}
     G_{\phi}(v)=\frac{\Gamma \left(\frac{d}{2}-1\right)}{(4\pi)^{\frac{d}{2}}}\left(1+\frac{a_0}{N}\right)f_{\rm bdry}\left(d-2+\hat\gamma_\phi,v\right)+\sum_{l=0}^{+\infty}\mu_l^2 f_{\rm bdry}\left(\hat{\Delta}_l,v\right)\,,
     \label{FulPropPhiPhi}
 \end{gather}
where $f_{\rm bdry}(\hat\Delta, v)$ denotes the boundary conformal block of scaling dimension $\hat\Delta$
\begin{gather}
    f_{\rm bdry}\left(\hat\Delta,v\right)=v ^{\hat\Delta} \, _2F_1\left(\hat\Delta,\hat\Delta-\frac{d}{2}+1;2 \hat\Delta-d+2;v\right)\,.
\end{gather}
In \eqref{FulPropPhiPhi},  $\hat\gamma_\phi\sim 1/N$ is the anomalous dimension of the leading boundary operator and $a_0$ is a constant irrelevant for our calculations. The subleading boundary operators are labeled by $l$, with dimensions $\hat\Delta_l$ and boundary OPE coefficients $\mu_l\sim 1/\sqrt{N}$. Since each boundary block is an eigenfunction of $\CD$, $\CD^2 G_\phi = G_\phi G_\sigma$ reduces to  algebraic constraints on the boundary data:
\begin{gather}
    \sum_l\left(\frac{(d-2\hat{\Delta}_l)(d-2-2\hat{\Delta}_l)}{4}-\frac{(d-2) (d-4)}{4}\right)^2\mu^2_l f_{\rm bdry}\left(\hat{\Delta}_l,v\right)=\frac{ \Gamma \left(\frac{d}{2}-1\right) }{(4 \pi)^{\frac{d}{2}}}\frac{v^{d-2}G_{\sigma}(v)}{(1-v)^{\frac{d}{2}-1}} \,,
\end{gather}
where we have used $f_{\rm bdry}(d-2, v) = v^{d-2}(1-v)^{1-d/2}$.
The small $v$ behavior of $G_\sigma$  fixes the possible dimensions appearing on the left-hand side to be $\hat{\Delta}_l=2d-2+l+O(1/N)$, with $l=0,1,2,\ldots$. We will see below that only even values of $l$ contribute.

It remains to determine the coefficients $\mu_l^2$. In \cite{Giombi:2020rmc}, these coefficients were obtained by expanding in powers of $\xi$ and solving iteratively order by order. Here we develop an alternative projection method, which is more systematic  and can be straightforwardly generalized to the GNY model. The starting point is the differential equation satisfied by the boundary blocks,
\begin{gather}
    \partial_v\left(v^{2-d}(1-v)^{\frac{d}{2}}\right)\partial_vf_{\rm bdry}\left(\Delta,v\right)=\Delta (\Delta-d+1)v^{-d}(1-v)^{\frac{d}{2}-1}f_{\rm bdry}\left(\Delta,v\right)\,.
\end{gather}
From this equation, one obtains the orthogonality relation, following  the standard Sturm-Liouville type argument
\begin{gather}
   \oint_C \frac{dv}{2\pi i} v^{-d}(1-v)^{\frac{d}{2}-1}f_{\rm bdry}\left(\Delta,v\right)f_{\rm bdry}\left(d-1-\Delta',v\right)=\delta_{\Delta,\Delta'}\,, \quad \quad \Delta-\Delta'=\mathbb{Z}\,,
   \label{OrthogonalityConditionO(N)}
\end{gather}
where $C$ is a contour encircling $v=0$ counterclockwise. This contour integral only makes sense when $\Delta-\Delta'=\mathbb{Z}$ because the leading small $v$ behavior of the integrand is $v^{\Delta-\Delta'-1}$. A noninteger $\Delta-\Delta'$ requires a branch cut that crosses the contour $C$.  In our case, this condition is automatically satisfied since the allowed boundary dimensions are $\hat\Delta_l = 2d -2+l+O(1/N)$.

Using the  orthogonality relation \eqref{OrthogonalityConditionO(N)}, we obtain the  following contour integral representation of the coefficient $\mu_l^2$ 
\begin{equation}
\begin{split}
    \mu^2_l
   &=Y_{l}\oint_C \frac{dv}{2\pi i} v^{-l-1} \, _2F_1(d-2,d;2d-4;v) \, _2F_1\left(2\!-\!\frac{3 d}{2}\!-\!l,1\!-\!d\!-\!l;4\!-\!3 d\!-\!2l;v\right)\,,\\
 Y_l&=\frac{1}{N}\frac{4^{1-d} (d-4)  \sin \left(\frac{\pi  d}{2}\right) \Gamma \left(\frac{d-1}{2}\right) \Gamma (d)}{ \pi ^{\frac{d}{2}+1}(d+l)^2 (2 d+l-3)^2 \Gamma \left(d-\frac{3}{2}\right)}~.
\end{split}
\end{equation}
By evaluating the residue at $v=0$, we find that $\mu^2_{2k+1}=0$ and
\begin{align}
   \mu^2_{2k}=Y_{2k}\sum _{n=0}^{2 k} \frac{(d-2)_n (d)_n}{n! (2 d-4)_n}\frac{ \left(-\frac{3 d}{2}-2 k+2\right)_{2 k-n} (-d-2 k+1)_{2 k-n}}{ (2 k-n)! (-3 d-4 k+4)_{2 k-n}}\equiv \frac{\Gamma \left(\frac{d}{2}-1\right)}{(4\pi)^{\frac{d}{2}}}\left(\mu^{O}_k\right)^2\,,
\end{align}
with $k=0,1,2,\ldots$. The coefficients $\left(\mu^{O}_k\right)^2$ are given by
\begin{gather}
    \left(\mu^{O}_k\right)^2=\frac{2^{-d-4 k+2} \sin \left(\frac{\pi  d}{2}\right) \Gamma \left(\frac{d-1}{2}\right) \Gamma \left(\frac{3 (d-1)}{2}+k\right) \Gamma \left(\frac{d}{2}+k\right) \Gamma (d+2 k)}{N\pi  d  (d+2 k) (2 d+2 k-3) k!\Gamma \left(\frac{d}{2}-2\right) \Gamma \left(\frac{d}{2}\right) \Gamma \left(d+k-\frac{1}{2}\right) \Gamma \left(\frac{3 (d-1)}{2}+2 k\right)}\,,
\end{gather}
and reproduce exactly eq.~(4.53) of \cite{Giombi:2020rmc}. Thus the boundary spectrum is $2d-2+2k+O(1/N)$.

The anomalous dimension $\hat\gamma_\phi$ can then be fixed by using the bulk OPE of $\phi^I\times \phi^J$. Precisely, the bulk OPE limit corresponds  to the $v \to 1$ limit, and the expansion  of $f_{\rm bdry}(\hat\Delta, v)$ in this limit contains two types of behavior: one proportional to $(1-v)^{1-d/2}$ and the other proportional to $(1-v)^0$. The first behavior corresponds to the identity operator in the bulk OPE limit and the second behavior corresponds to a $\Delta = d-2$ bulk operator. However, in the large $N$ limit, there is no dimension $d-2$ operator in the bulk OPE of $\phi^I\times \phi^J$. In \eqref{FulPropPhiPhi}, requiring the $(1-v)^0$ term to vanish at order $1/N$ \footnote{The leading boundary conformal block $f_{\rm bdry}(d-2,v)$ does not contain the $(1-v)^0$ term in the bulk OPE expansions, and hence the coefficient $a_0$ does not enter the calculation.}, we get 
\begin{gather}
    \hat\gamma_\phi=-\frac{\Gamma\left(\frac{d}{2}-1\right)}{\Gamma(d-2)}\sum \limits_{k=0}^{+\infty}\left(\mu^{O}_k\right)^2\frac{\Gamma\left(3d-2+4k\right)}{\Gamma(d+2k)\Gamma\left(\frac{3d}{2}-1+2k\right)}\,.
\end{gather}
Recently, this representation was further analyzed in \cite{Ohno:2025qfv}, where it was shown that the sum can be rewritten as a generalized hypergeometric series
\begin{gather}
    \hat\gamma_\phi
    =-\frac{2^{d+1} (3 d-3) \sin \left(\frac{\pi  d}{2}\right) \Gamma \left(\frac{3d-3}{2}\right)}{N\pi  d^2 (2 d-3)  \Gamma \left(\frac{d}{2}-2\right) \Gamma \left(d-\frac{1}{2}\right)}\sum \limits_{k=0}^{+\infty}\frac{\left(\frac{3 d+1}{4} \right)_k\left(\frac{d}{2}\right)_k\left(d-\frac{3}{2}\right)_k \left(\frac{3d-3}{2}\right)_k \left(\frac{d}{2}\right)_k  }{k! \left(\frac{d}{2}+1\right)_k \left(d-\frac{1}{2}\right)_k  \left(\frac{3 d-3}{4} \right)_k\left(d-\frac{1}{2}\right)_k}\nonumber\\
    =-\frac{2^{d+1} (3 d-3) \sin \left(\frac{\pi  d}{2}\right) \Gamma \left(\frac{3 (d-1)}{2}\right) }{N\pi  d^2 (2 d-3)  \Gamma \left(\frac{d}{2}-2\right) \Gamma \left(d-\frac{1}{2}\right)}{}_5F_4\!\left(\begin{matrix}\frac{3d+1}{4},\; \frac{d}{2},\; d-\frac{3}{2},\; \frac{3d-3}{2},\; \frac{d}{2}\\[2pt] \frac{d}{2}+1,\; d-\frac{1}{2},\; \frac{3d-3}{4},\; d-\frac{1}{2}\end{matrix}\;\bigg|\; 1\right)\,.
\end{gather}
The hypergeometric function in the last line belongs to the class of
${}_5F_4(1)$ series that admits a closed-form evaluation in terms of Gamma
functions. Applying the identity used in \cite{Ohno:2025qfv} and simplifying
the result, one finds
\begin{gather}
    \hat\gamma_\phi=
    \frac{1}{N}
    \frac{(4-d)\Gamma(2d-3)}
    {d\Gamma(d-2)\Gamma(d-1)}\,,
\end{gather}
in agreement with \cite{Ohno:1983lma}.

\subsubsection{Fermion boundary anomalous dimension in the GNY model}

In order to generalize the equation-of-motion method to the GNY model, we notice that for a fixed boundary dimension $\hat \Delta$,
there are two inequivalent spinor boundary conformal blocks $\CW_\pm(\hat\Delta,\xi)$, satisfying, up to contact terms 
\begin{align}
\Dn_1\,  \CW_\pm(\hat\Delta,\xi)=   -\CW_\pm(\hat\Delta,\xi)\, \lDn_2= \pm\left(\hat\Delta- \frac{d-1}{2}\right)\, \CW_\pm(\hat\Delta, \xi)~.
\end{align}
Each $\CW_\pm$  carries two tensor structures \cite{Herzog:2022jlx}:
\begin{align}\label{Walpha}
\CW_\alpha(\hat\Delta, \xi) = \frac{1}{2\sqrt{z_1 z_2}}\left[\frac{\slashed{x}_{12}}{\sqrt{\xi}}\, \hat f_\alpha(\hat\Delta, \xi) - \frac{\gamma_0\slashed{\bar x}_{12}}{\sqrt{\xi+1}}\, \hat g_\alpha(\hat\Delta, \xi)\right], \quad \alpha =\pm ~,
\end{align}
with the component functions being 
\begin{align}
\hat f_\alpha(\hat\Delta, \xi) &= \xi^{-\hat\Delta}\, \F\!\left(\hat\Delta+\tfrac{1}{2},\; \hat\Delta-\tfrac{d-1}{2};\; 2\hat\Delta-d+2;\; -\tfrac{1}{\xi}\right)~,\label{fhat_xi}\\[4pt]
\hat g_\alpha(\hat\Delta, \xi) &= -\alpha\sqrt{\tfrac{\xi+1}{\xi}}\;\xi^{-\hat\Delta}\, \F\!\left(\hat\Delta+\tfrac{1}{2},\; \hat\Delta-\tfrac{d-3}{2};\; 2\hat\Delta-d+2;\; -\tfrac{1}{\xi}\right)~.\label{ghat_xi}
\end{align}
For later convenience, it is useful to express the component functions in terms of $v = \frac{1}{\xi+1}$, using the relation 
$\F\left(a,b;c;-\tfrac{1}{\xi}\right) = (1-v)^{a}\,\F(a,c-b;c;v)$:
\begin{align}
\hat f_\alpha(\hat\Delta, v) &= \sqrt{1-v}\; v^{\hat\Delta}\, \F\!\left(\hat\Delta - \tfrac{d-3}{2},\; \hat\Delta+\tfrac{1}{2};\; 2\hat\Delta - d + 2;\; v\right)~,\label{fhat}\\[4pt]
\hat g_\alpha(\hat\Delta, v) &= -\alpha\, v^{\hat\Delta}\, \F\!\left(\hat\Delta - \tfrac{d-1}{2},\; \hat\Delta+\tfrac{1}{2};\; 2\hat\Delta - d + 2;\; v\right)~.\label{ghat}
\end{align}
In this new variable, the boundary OPE limit corresponds to $v=0$.

In the large $N$ limit, as discussed around \eqref{saddlenormal}, the normal boundary phase corresponds to the saddle point $\sigma_\star = \frac{d}{2}-1 \equiv \mu_\star$.  At this fixed point, the free fermion propagator \eqref{GPsi^nor} is  given by the boundary conformal block $\CW_-$ with dimension $\hat{\Delta}=d-\tfrac{3}{2}$:
\begin{align}\label{Gpsi0}
G_\Psi^{\rm nor}(x_1,x_2) = \CN_\Psi\, \CW_-\left(d-\tfrac{3}{2}, \xi\right)~, \qquad \CN_\Psi = -\frac{\Gamma(\frac{d}{2})}{(4\pi)^{d/2}}~.
\end{align}
The bulk Yukawa interaction leads to an anomalous dimension $\hat{\gamma}_\Psi$ at the order of $1/N$.

Let $G_\Psi(x_1,x_2)$ denote the fermion two-point function, including the leading $1/N$ correction. Acting with the Dirac operator on both insertions and using the equations of motion, we obtain, at leading order in the large $N$ expansion,
\begin{align}\label{GNY_EOM}
(\Dn_1 + \mu_\star)\, G_\Psi(x_1,x_2)\,(-\lDn_2 + \mu_\star) = G^{\rm nor}_{\sigma}(\xi)\, G_\Psi^{\rm nor}(x_1,x_2)~.
\end{align}
where in deriving this relation, we used that the four-point function $\langle \sigma(x_1)\Psi_I(x_1)\sigma(x_2)\bar\Psi^J(x_2)\rangle$ factorizes at leading order in $1/N$. The scalar two-point function $G^{\rm nor}_{\sigma}(\xi)$ \eqref{Gsigmanorm} can be written in $v$ coordinates as
\begin{align}\label{Bsigma}
G^{\rm nor}_{\sigma}(\xi) = B_\sigma\, v^d\, \F(d, d-1; 2d-2; v)\,, \qquad  B_\sigma = -\frac{1}{Nc_d}\frac{2^{2d-5}(d-2)\Gamma\!\left(\frac{d-1}{2}\right)^2\Gamma(d)}{\pi\,\Gamma\!\left(\frac{d}{2}\right)\Gamma\!\left(1-\frac{d}{2}\right)\Gamma(2d-2)}\,.
\end{align}
The most general boundary block expansion of $G_\Psi(x_1,x_2)$ takes the form 
\begin{align}\label{an}
G_\Psi(x_1,x_2) &= \CN_\Psi\left[\left(1+\frac{a_0}{N}\right)\CW_-\left(d-\tfrac{3}{2}+\hat{\gamma}_\Psi, \xi\right)+\sum \limits_{l=0}^{+\infty}\sum_{\alpha =\pm}\kappa_{l,\alpha}^2 \CW_{\alpha}(\hat\Delta_l,\xi)\right]\nonumber\\
&\equiv \frac{\CN_\Psi}{2\sqrt{z_1 z_2}}\left[\frac{\slashed{x}_{12}}{\sqrt{\xi}}F(\xi) - \frac{\gamma_0\slashed{\bar x}_{12}}{\sqrt{\xi+1}}G(\xi)\right]~,
\end{align}
where $\kappa_{l,\alpha}^2\propto 1/N$.
Substituting \eqref{an} into \eqref{GNY_EOM}, the first block does not contribute at order $1/N$, and each of the remaining blocks picks up a factor $(\mu_\star + \alpha\hat{\mu}_l)^2$: 
\begin{align}\label{EOM_expanded}
\CN_\Psi\sum\limits_{l=0}^{+\infty}(\mu_\star + \hat{\mu}_l)^2\,\kappa^2_{l,+}\,\CW_+(\hat\Delta_l,\xi) + \CN_\Psi\sum\limits_{l=0}^{+\infty}(\mu_\star - \hat{\mu}_l)^2\,\kappa^2_{l,-}\,\CW_-(\hat\Delta_l,\xi) = G^{\rm nor}_{\sigma}\, G_\Psi^{\rm nor}~,
\end{align}
where $\hat{\mu}_l = \hat\Delta_l - \frac{d-1}{2}$.
Since the two tensor structures in $\CW_\alpha(\hat\Delta, \xi)$ are independent, \eqref{EOM_expanded} naturally splits into two equations
\begin{align}
\sum_{l}\Big[(\mu_\star{+}\hat{\mu}_l)^2\kappa^2_{l,+} + (\mu_\star{-}\hat{\mu}_l)^2\kappa^2_{l,-}\Big]\hat f_+(\hat\Delta_l,v) &= B_\sigma(1{-}v)^{\frac{1-d}{2}} v^{2d-\frac{3}{2}}\, \F(d,d{-}1;2d{-}2;v)~,\label{EOM_f}\\[4pt]
\sum_{l}\Big[(\mu_\star{+}\hat{\mu}_l)^2\kappa^2_{l,+} - (\mu_\star{-}\hat{\mu}_l)^2\kappa^2_{l,-}\Big]\hat g_+(\hat\Delta_l,v) &= B_\sigma(1{-}v)^{1-\frac{d}{2}} v^{2d-\frac{3}{2}}\, \F(d,d{-}1;2d{-}2;v)~.\label{EOM_g}
\end{align}
where we have used that  $\hat f_- = \hat f_+$ and $\hat g_- = -\hat g_+$.
The small $v$ behavior of \eqref{EOM_f} and \eqref{EOM_g} indicates boundary fermions of scaling dimension $\hat\Delta_l = 2d - \frac{3}{2} + l$ with $l = 0, 1, 2, \ldots$.

The coefficients $\kappa^2_{l,\pm}$ can be determined analytically because the functions entering the fermionic boundary blocks satisfy simple second-order differential equations. To make this structure manifest, we define
\begin{align}
\tilde f(\hat\Delta,v) = (1-v)^{-1/2}\hat f_+(\hat\Delta,v)\,,\qquad
\tilde g(\hat\Delta,v) = v^{-(d-1)/2}\hat g_+(\hat\Delta,v)\,.
\end{align}
One then finds that $\tilde g$ and $\tilde f$ satisfy
\begin{equation}\label{gtilde_eigen}
    \begin{split}
       &\partial_v\!\Big(v(1-v)^{d/2}\,\partial_v\tilde g\Big) = \frac{(2\hat\Delta - d + 1)^2}{4}\,\frac{(1-v)^{d/2-1}}{v}\,\tilde g~,\\
&\partial_v\!\Big(v^{2-d}(1-v)^{d/2+1}\,\partial_v\tilde f\Big) + \frac{d-3}{4}\,v^{1-d}(1-v)^{d/2}\,\tilde f = \hat\Delta(\hat\Delta {-} d {+} 1)\,v^{-d}(1-v)^{d/2}\,\tilde f~. 
    \end{split}
\end{equation}
These equations imply the orthogonality relations (similarly to \eqref{OrthogonalityConditionO(N)})
\begin{align}
\oint_C \frac{dv}{2\pi i}\, v^{-d}(1-v)^{d/2}\, \tilde f(\hat\Delta, v)\, \tilde f(d{-}1{-}\hat\Delta', v) &= \delta_{\hat\Delta, \hat\Delta'}~,\label{ortho_f}\\[4pt]
\oint_C \frac{dv}{2\pi i}\, \frac{(1-v)^{d/2-1}}{v}\, \tilde g(\hat\Delta, v)\, \tilde g(d{-}1{-}\hat\Delta', v) &= \delta_{\hat\Delta, \hat\Delta'}~,\label{ortho_g}
\end{align}
provided $\hat\Delta-\hat\Delta' \in \mathbb{Z}$, where $C$ encircles $v=0$ counterclockwise. 

We now use the orthogonality relations to extract the OPE coefficients from \eqref{EOM_f}--\eqref{EOM_g} using similar logic as around \eqref{OrthogonalityConditionO(N)}. The contour projections of \eqref{EOM_f} and \eqref{EOM_g} reduce to terminating ${}_3F_2$ functions. The only difference between the two projections is in one lower parameter of the resulting terminating ${}_3F_2$: \eqref{EOM_f} gives $\frac{3d}{2}-1$, whereas \eqref{EOM_g} gives $\frac{3d}{2}$. These two hypergeometric functions are related by the identity\footnote{This identity follows from transformation formula for
${}_3F_2(1)$ \cite{Bailey1935} (see in particular Ch.~III, Sec.~3.8, Eq.~1). We apply it with
\(a=-k\), \(b=3d+k-2\), \(c=d-1\), \(e=\frac{3d}{2}-1\), and \(f=2d-2\).}
\begin{align}\label{3F2_relation}
{}_3F_2\!\left(\begin{matrix} {-}k,\; 3d{+}k{-}2,\;d{-}1\\[2pt] \tfrac{3d}{2}-1,\; 2d{-}2\end{matrix}\;\bigg|\; 1\right) = \frac{(-1)^k(3d+2k-2)}{3d-2}\, {}_3F_2\!\left(\begin{matrix}{-}k,\; 3d{+}k{-}2,\;d{-}1\\[2pt] \tfrac{3d}{2},\; 2d{-}2\end{matrix}\;\bigg|\; 1\right)~.
\end{align}
It is therefore convenient to express both projections in terms of
\begin{align}\label{Tk_3F2}
T_k \equiv -B_\sigma\,\frac{(-\frac{3d}{2}-k+1)_k\,(-d-k+1)_k}{k!\,(-3d-2k+3)_k}\; 
 {}_3F_2\!\left(\begin{matrix}{-}k,\; 3d{+}k{-}2,\;d{-}1\\[2pt] \tfrac{3d}{2},\; 2d{-}2\end{matrix}\;\bigg|\; 1\right)~.
\end{align}
With this notation, the projected equations at fixed $k$ become
\begin{align}
(\mu_\star{+}\hat{\mu}_k)^2\kappa^2_{k,+} + (\mu_\star{-}\hat{\mu}_k)^2\kappa^2_{k,-} &= -(-1)^k T_k~,\label{system_plus}\\[2pt]
(\mu_\star{+}\hat{\mu}_k)^2\kappa^2_{k,+} - (\mu_\star{-}\hat{\mu}_k)^2\kappa^2_{k,-} &= \phantom{-}T_k~.\label{system_minus}
\end{align}
The two equations imply a simple parity selection rule: $\kappa_{k,+}$ vanishes for even $k$, whereas $\kappa_{k,-}$ vanishes for odd $k$. The nonzero coefficients are therefore
\begin{align}\label{kappas}
\kappa^2_{2n+1,+} = \frac{T_{2n+1}}{(\mu_\star+\mu_{2n+1})^2}~, \qquad \kappa^2_{2n,-} = -\frac{T_{2n}}{(\mu_\star-\mu_{2n})^2}~, \qquad n = 0, 1, 2, \ldots\,.
\end{align}
One may check that these coefficients are positive for $1<d<4$.

To obtain a closed-form expression for the ${}_3F_2$ function appearing in \eqref{Tk_3F2}, we first recall that Watson's summation theorem \cite{Bailey1935} gives:
\begin{align}
 {}_3F_2\!\left(\begin{matrix}u,\; v,\; c\\[2pt] \tfrac{u+v+1}{2},\; 2c\end{matrix}\;\bigg|\; 1\right)
=
\frac{\Gamma(\frac12)\Gamma(\frac{u+v+1}{2})\Gamma(c+\frac12)\Gamma(c-\frac{u+v}{2}+\frac12)}
{\Gamma(\frac{u+1}{2})\Gamma(\frac{v+1}{2})\Gamma(c-\frac u2+\frac12)\Gamma(c-\frac v2+\frac12)}\,.
\label{eq:Watson}
\end{align}
Next, the elementary identity $a(a+1)_j(b)_j-b(a)_j(b+1)_j=(a-b)(a)_j(b)_j$, which, when applied term by term to the defining series of ${}_3F_2$, implies
\begin{align}
a\, {}_3F_2\!\left(\begin{matrix}a+1,\; b,\; c\\[2pt] \tfrac{a+b+2}{2},\; 2c\end{matrix}\;\bigg|\; 1\right)-b\, {}_3F_2\!\left(\begin{matrix}a,\; b+1,\; c\\[2pt] \tfrac{a+b+2}{2},\; 2c\end{matrix}\;\bigg|\; 1\right)=(a-b)\,{}_3F_2\!\left(\begin{matrix}a,\; b,\; c\\[2pt] \tfrac{a+b+2}{2},\; 2c\end{matrix}\;\bigg|\; 1\right)\,.
\label{eq:contigF}
\end{align}
The hypergeometric function appearing in $T_k$ is of precisely of the form in RHS in \eqref{eq:contigF}, upon identifying $a=-k$, $b=3d+k-2$, and $c=d-1$. Combining \eqref{eq:Watson} with \eqref{eq:contigF}, we then obtain
\begin{align}
{}_3F_2\!\left(\begin{matrix}d{-}1,\; {-}k,\; 3d{+}k{-}2\\[2pt] \tfrac{3d}{2},\; 2d{-}2\end{matrix}\;\bigg|\; 1\right)
=
\begin{cases}
\displaystyle
\frac{2\,\Gamma(d{-}\frac{1}{2})\,\Gamma(\frac{3d}{2})\,\Gamma(n{+}\frac{1}{2})\,\Gamma(\frac{d}{2}{+}n{+}1)}
{\sqrt{\pi}\,(3d{+}4n{-}2)\,\Gamma(\frac{d}{2}{+}1)\,\Gamma(d{+}n{-}\frac{1}{2})\,\Gamma(\frac{3d}{2}{+}n{-}1)},
& k=2n\,, \\[1.2em]
\displaystyle
\frac{2\Gamma(d{-}\frac{1}{2})\,\Gamma(\frac{3d}{2})\,\Gamma(n{+}\frac{3}{2})\,\Gamma(\frac{d}{2}{+}n{+}1)}
{\sqrt{\pi}\,(3d{+}4n)\,\Gamma(\frac{d}{2}{+}1)\,\Gamma(d{+}n{-}\frac{1}{2})\,\Gamma(\frac{3d}{2}{+}n)},
& k=2n+1\,.
\end{cases}
\label{3F2_closed}
\end{align}

Having determined all the $\kappa$'s, we can now extract the anomalous dimension $\hat{\gamma}_\Psi$. 
Recall the full propagator \eqref{an}. To fix $\hat{\gamma}_\Psi$, we study the bulk OPE limit $\xi\to 0$. In this limit, the component $\hat g_+$ has the expansion
\begin{equation}\label{ghat_expansion}
\begin{split}
\hat g_+(\hat\Delta, \xi) = (4\xi)^{1-d/2}&\bigg(-\frac{\Gamma(\frac{d}{2}-1)\,\Gamma(2\hat\Delta-d+2)}{\Gamma(\hat\Delta+\frac{1}{2})\,\Gamma(\hat\Delta - \frac{d-1}{2})} + O(\xi)\bigg)\\
&+ \bigg(-\frac{\Gamma(1-\frac{d}{2})\,\Gamma(2\hat\Delta-d+2)}{\Gamma(\hat\Delta-d+\frac{3}{2})\,\Gamma(\hat\Delta - \frac{d-3}{2})} + O(\xi)\bigg)~.
\end{split}
\end{equation}
The first line contains a power-law divergence $\xi^{1-d/2}$, belonging to the identity channel in the bulk OPE. The constant term in the second line would correspond to a bulk scalar of dimension $2\Delta_{\Psi}=d-1$, identified in the free theory with the composite operator $\bar\Psi\Psi$. In  the large $N$ GN CFT, the singlet scalar appearing in the bulk OPE is instead $\sigma$ with scaling dimension $\Delta_\sigma=1+O\left(\frac{1}{N}\right)$. Therefore, consistency with the bulk OPE requires the total coefficient of the term $\xi^0$ to vanish after summing all contributions at order $1/N$.

To implement this constraint, we collect the constant terms in the $\xi\to 0$ expansion of the component $G(\xi)$ (see \eqref{an}) of the full propagator. \footnote{The same procedure could be implemented for $F(\xi)$ which leads to the same result.} Using $\hat g_- = -\hat g_+$, the leading block contributes $\hat{\gamma}_\Psi$ times the derivative of the block with respect to $\hat\Delta$, and each double-trace block contributes $(\kappa^2_{l,+} - \kappa^2_{l,-})$ times its constant term. Setting their sum to zero:
\begin{align}\label{gammaO_sum}
\hat{\gamma}_\Psi = \sum_{l=0}^{\infty} \left(\kappa^2_{l,+}-\kappa^2_{l,-}\right)\frac{4^{d+l}\,\Gamma(\frac{3d}{2}+l-\frac{1}{2})}{\Gamma(\frac{d-1}{2})\,\Gamma(d+l)}~.
\end{align}
Substituting the OPE coefficients \eqref{kappas} and the closed forms \eqref{3F2_closed}, the even ($l=2n$, $\alpha=-$) and odd ($l=2n+1$, $\alpha=+$) contributions combine into a single sum over $n$
\begin{align}\label{gammaO_explicit_sum}
\hat{\gamma}_\Psi &= \frac{B_\sigma 2^{3d-1}\Gamma(d-\frac{1}{2})}{\sqrt{\pi}\,\Gamma(d\pm 1)}\sum_{n=0}^{\infty}\left(\frac{1}{(2d+2n-1)^2}-\frac{1}{(d+2n)^2}\right)\frac{\Gamma(\frac{d}{2}+n+1)\,\Gamma(\frac{3d}{2}+n-\frac{1}{2})}{\Gamma(n+1)\,\Gamma(d+n-\frac{1}{2})}~.
\end{align}
This sum can be performed in closed form. The key step is the partial fraction decomposition
\begin{align}\label{partial_frac}
&\left(\frac{1}{(2d+2n-1)^2}-\frac{1}{(d+2n)^2}\right)\frac{\Gamma(\frac{d}{2}+n+1)\,\Gamma(\frac{3d}{2}+n-\frac{1}{2})}{\Gamma(n+1)\,\Gamma(d+n-\frac{1}{2})}\nonumber\\
&=-\frac{(d-1) \Gamma \left(\frac{d}{2}+1\right) \Gamma \left(\frac{3 d}{2}+\frac{1}{2}\right)}{d^2 (2 d-1) \Gamma \left(d+\frac{1}{2}\right)}\frac{\left(\frac{3 d-1}{2} \right)_n \left(1+\frac{3 d-1}{4}\right)_n \left(\frac{d}{2}\right)_n \left(\frac{d}{2}\right)_n \left(d-\frac{1}{2}\right)_n}{n! \left(\frac{3 d-1}{4} \right)_n \left(d+\frac{1}{2}\right)_n \left(d+\frac{1}{2}\right)_n \left(\frac{d}{2}+1\right)_n}\,.
\end{align}
After this rewriting, the summand takes the form $\frac{(a_1)_n\cdots(a_5)_n}{n!(b_1)_n\cdots(b_4)_n}$, which is by definition a well-poised ${}_5F_4$ at unit argument:
\begin{align}\label{gammaO_resum}
\hat{\gamma}_\Psi &\propto {}_5F_4\!\left(\begin{matrix}\frac{3d-1}{2},\; 1+\frac{3d-1}{4},\; \frac{d}{2},\; \frac{d}{2},\; d-\frac{1}{2}\\[2pt] \frac{3d-1}{4},\; d+\frac{1}{2},\; d+\frac{1}{2},\; \frac{d}{2}+1\end{matrix}\;\bigg|\; 1\right)~.
\end{align}
This ${}_5F_4$ has a special structure: the numerator contains both $\alpha = \frac{3d-1}{2}$ and $1+\frac{\alpha}{2} = \frac{3d+1}{4}$, while the denominator contains $\frac{\alpha}{2} = \frac{3d-1}{4}$. This is the hallmark of a \emph{well-poised} hypergeometric series, for which the Dougall--Dixon summation formula provides a closed-form evaluation in terms of Gamma functions.\footnote{See in particular Ch.~IV, Sec.~4.4, Eq.~1 in \cite{Bailey1935}, and the recent discussion in \cite{Ohno:2025qfv}.} After simplification we get:
\begin{align}\label{gammaO_final}
\hat{\gamma}_\Psi = \frac{1}{Nc_d}\,\frac{2^{2d-3}\,\Gamma(d-\frac{1}{2})}{\sqrt{\pi}\, d\,\Gamma(d-2)}~.
\end{align}
Let us perform several consistency checks. In $d=2+\epsilon$, our result gives $\hat\Delta_{(1/2)}=\frac{1+2\epsilon}{2}+\frac{1}{4N} (\epsilon +O(\epsilon^2))$. This agrees with the known $\epsilon$-expansion result of \cite{Giombi:2021cnr}, where the scaling dimension of the leading boundary fermion was found to be $\hat{\Delta}^{(d=2+\epsilon)}_{(1/2)}=\frac{1}{2}+\frac{4N-3}{4(N-1)}\epsilon$. Expanding this expression at large $N$, we find perfect agreement with our result.  Finally, expanding \eqref{gammaO_final} around $d=4-\epsilon$, we find
\begin{align}\label{4e}
    \hat\Delta_{(1/2)} = \frac{3-\epsilon}{2}+\frac{1}{N}\left(\frac{15}{4} - \frac{109}{16}\,\epsilon + \left(\frac{5\pi^2}{8} - \frac{29}{64}\right)\epsilon^2 + O(\epsilon^3)\right)+O\left(1/N^2\right)~.
\end{align}
The $\epsilon$-expansion in \cite{Giombi:2021cnr} gives
$\hat\Delta^{(d=4-\epsilon)}_{(1/2)} = \frac{3}{2} + \sqrt{\frac{36}{\sqrt{4N^2+132N+9}-2N+3}}+\CO(\epsilon)$, whose large $N$ expansion reproduces the term $\frac{15}{4N}\epsilon^0$ in \eqref{4e}.

Finally, let us note that in $3d$, \eqref{gammaO_final} gives $\hat{\gamma}_\Psi = 1/N$, and hence we get that
\begin{gather}
    \hat\Delta^{(d=3)}_{(1/2)} = \frac{3}{2} + \frac{1}{N} + O(1/N^2)\,.
\end{gather}

\section{The $\epsilon$ expansion}\label{Sec:Eps_Expansion}
In this section we compute corrections to the one-point function $\langle s \rangle$ as well as the free energy in the normal universality class to order $\epsilon$.

In contrast to the Wilson-Fisher case, there is an additional technical complication due to the existence of two couplings. 
For example, as reviewed in section  \ref{normalreview}, the bare fermion mass is not a constant but proportional to  $\frac{g_{1,0}}{\sqrt{g_{2,0}}}$.  As a result, some care is required when rewriting bare quantities in terms of renormalized couplings. Notice that, at  the fixed point, the two couplings scale differently, namely $g_{1,\star}\sim \sqrt{\epsilon}$ and $g_{2,\star}\sim \epsilon$, c.f. \eqref{fp}. A convenient way to keep track of the order of perturbation is to introduce an auxiliary parameter $\alpha$ and make the replacement $g_1\to \sqrt{\alpha}\, g_1$ and $g_2\to \alpha\, g_2$ for renormalized couplings. Expanding in $\alpha$ then automatically organizes the result in powers of $\sqrt{\epsilon}$. After truncating at the required order, we set $\alpha=1$.

\subsection{One-point function of $s$}\label{SubSec:1pfInEpsExpansion}
\begin{figure}[t]
\centering
\begin{tikzpicture} 
\draw[color=black,dashed] (0,0) to (0,1); 
\draw[color=black,dashed] (0,1.5) circle [radius=0.5]; 
\node at (0,-0.5) {$\mathcal{T}^{}_{1}$};
\draw (0,1)node[vertex]{} ;
\draw (0,0)node[vertex]{} ;
\end{tikzpicture} 
\quad
\begin{tikzpicture} 
\draw[color=black,dashed] (0,0) to (0,1); 
\draw[color=black] (0,1.5) circle [radius=0.5]; 
\node at (0,-0.5) {$\mathcal{T}^{}_{2}$};
\draw (0,1)node[vertex]{} ;
\draw (0,0)node[vertex]{} ;
\end{tikzpicture} 
\caption{Leading contributions to the tadpole $\langle t \rangle$. Dashed lines denote scalar propagators, while solid lines denote fermion propagators.}
\label{Tadpole}
\end{figure}
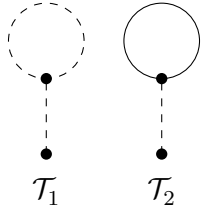
In this subsection, we compute the first correction to $ \langle s_0 \rangle = s_{\star,0}+\langle t\rangle$, and compare the result with the large-$N$ prediction of \cite{Giombi:2021cnr}. At this order, $\langle t\rangle$ receives contributions from the two tadpole diagrams shown in Figure \ref{Tadpole}:
\begin{gather}
    \langle t\rangle=\mathcal{T}_1+\mathcal{T}_2+O\left(g^{n_1}_{1,0}\left(g_{2,0}\right)^{\frac{n_2}{2}}\right)\big|_{n_1+n_2=3}\,.
    \label{1pftAsSumOfT1PT2}
\end{gather}
Using \eqref{Gpsi(x,x)} and \eqref{Gt(x,x)}, together with the identity 
\begin{gather}
    \int_x G_t(x,y)=\frac{1}{\hat{\Delta}_t(\hat{\Delta}_t-d+1)}\,,
\end{gather}
we find
\begin{align}
    \mathcal{T}_1
    &= -\frac{\sqrt{g_{2,0}} }{2}\sqrt{\frac{3d(d-2)}{2}} \, G_t(1)\int_x G_t(x,y)
    =-\sqrt{g_{2,0}}\sqrt{\frac{3}{2d(d-2)}}\,
    \frac{\Gamma\left(1-\frac{d}{2}\right)\Gamma(\hat{\Delta}_t)}{(4\pi)^{d/2}\Gamma(\hat{\Delta}_t-d+2)}\,,\nonumber\\
    \mathcal{T}_2
    &= g_{1,0}\langle \bar{\Psi}_0 \Psi_{0}\rangle\int_x G_t(x,y)
    =-g_{1,0}\frac{4N\,\Gamma\left(-\frac{d}{2}\right)}{(2\pi)^{d/2}(d-2)}
    \frac{\Gamma\left(\frac{d}{2}+\mu_0\right)}{\Gamma\left(1-\frac{d}{2}+\mu_0\right)}\,,
\end{align}
where the first contribution, $\mathcal{T}_1$, was already computed in \cite{Giombi:2025pxx}.

Unlike in the O$(N)$ case, wavefunction renormalization contributes already at this order. Indeed, in both theories it starts at quadratic order in the couplings, but in the present case this is order $g_1^2\sim \epsilon$ at the fixed point. Therefore one must include the leading wavefunction renormalization
$Z_s=1-\frac{N g_1^2}{4\pi^2\epsilon}$. 
Combining all contributions and expanding systematically in terms of renormalized couplings, following the prescription discussed in the beginning of Section ~\ref{Sec:Eps_Expansion}, we obtain
\begin{align}\label{s1}
    \langle s \rangle
    &=\frac{s_{\star,0}+\langle t \rangle}{\sqrt{Z_{s}}}
    =\frac{\sqrt{3} (8-3 \epsilon )}{4 \sqrt{g_2 }}
    +\sqrt{g_2} \frac{\sqrt{3} (6 \Upsilon -11)}{64 \pi ^2}
    +\frac{g_{1} N (36 g^2_1-g_2 )}{16 \pi ^2 g_2 }\nonumber\\
    &-\frac{\sqrt{3} g^2_{1} N \big(24 g^2_1 -5g_2 \big)}{16 \pi ^2 g_{2} ^{3/2}}
    +\frac{\sqrt{3} g^2_1 N (12 g^2_1-g_2 )}{8 \pi ^2 g_2 ^{3/2}}\left(2H_{ \frac{2\sqrt{3}g_1}{\sqrt{g_2} }-2}-\Upsilon\right)\,,
\end{align}
where $H_s$ denotes the harmonic numbers and 
\begin{gather}
    \Upsilon=\gamma_E+\log{(4\pi)}\,.
    \label{UpsilonDefinition}
\end{gather}
 The first two terms in \eqref{s1}  coincide with the result for the O$(N)$ model in \cite{Giombi:2025pxx} upon setting $N=1$. Evaluating this expression at the fixed point, we find
\begin{align}\label{sfpN}
  \langle s \rangle_{\rm f.p.}& =\frac{3 }{ \pi  S_N}\sqrt{N+\frac{3}{2}}\frac{1}{\sqrt{\epsilon}}+\sqrt{\epsilon}\Bigg(\frac{9 \Upsilon }{4 \pi S_N \sqrt{4N+6} }+\frac{3  \left(S_N^2-6\right)}{4 \pi  \sqrt{4N+6} S_N}H_{-2+\frac{6}{S_N}}+\frac{8N+3 \left(S_N^2-6\right)}{16 \pi  \sqrt{4N+6}}\nonumber\\
  &+\frac{\left(78 N^2-166 N-237\right) S_N^2-90 \left(90 N^2+146 N+9\right)}{24 \sqrt{2} \pi  (2N+3)^{3/2} S_N \left(S_N^2+2N-3\right)}\Bigg)\,,
\end{align}
where
\begin{gather}
    S_N=\sqrt{\sqrt{4 N^2+132 N+9}-2 N+3}\,.
    \label{SNDefinition}
\end{gather}
In the large-$N$ limit, \eqref{sfpN} implies $ \lim\limits_{N \to \infty} g_{1,\star} \langle s \rangle_{\rm f.p.}=1-\frac{\epsilon}{2}$,
where we have used the fixed point \eqref{fp}.
It is in agreement with the large-$N$ calculation of \cite{Giombi:2021cnr}.

To compare with the standard BCFT literature, we normalize the one-point function of $s$ by the bulk two-point function coefficient. We therefore define
\begin{gather}
    a^{\rm nor}_s=\frac{2^{\Delta_s}}{\mathcal{N}_s}\,\langle s\rangle_{\rm f.p.}\, .
\end{gather}
Here $\mathcal{N}_s$ is the flat-space normalization of the operator $s$, and $\Delta_s$ is its scaling dimension, i.e. $\langle s(x) s(0)\rangle = \tfrac{\CN_s^{\,2}}{x^{2\Delta_s}}\,$
 at the IR fixed point of the GNY model.
 Their explicit expressions to order $\epsilon$ are \cite{Barrat:2023ivo}
\begin{gather}
    \mathcal{N}_{s}^{\,2}=\frac{1}{4 \pi ^2}+\frac{3 \log (\pi e^{\gamma_E})-4N}{8 \pi ^2(2N+3)} \epsilon\,, \, \quad 
    \Delta_s=1-\frac{3\epsilon}{2(2N+3)}\, .
\end{gather}
With this normalization, one finds
\begin{align}\label{asgen}
    a^{\rm nor}_s&=\frac{6 \sqrt{4N+6}}{ S_N}\frac{1}{\sqrt{\epsilon}}+\sqrt{\epsilon}\bigg(\frac{3  \left(S_N^2-6\right)}{\sqrt{4N+6} S_N}H_{-2+\frac{6}{S_N}}+\frac{8N(S_N+6)+3 S_N \left(S_N^2-6\right)}{4 \sqrt{4N+6} S_N}\nonumber\\
    &+\frac{\left(78N^2-166N-237\right) S_N^2-90 \left(90 N^2+146 N+9\right)}{6 \sqrt{2} (2N+3)^{3/2} S_N \left(S_N^2+2N-3\right)}\bigg)~.
\end{align}
For some small $N$, we give the explicit expression
\begin{equation}\label{assusy}
    \begin{split}
         &N=\frac{1}{4}: \quad  a^{\rm nor}_s = 2\sqrt{\frac{7}{\epsilon}}\left(1-\frac{27}{28}\epsilon+O(\epsilon^2)\right)\,,\\   
      &N=\frac{1}{2}: \quad  a^{\rm nor}_s = \frac{1}{\sqrt{\epsilon}}\left(5.18374-4.74556 \epsilon+O(\epsilon^2)\right)\,,\\
   &N=1: \quad  a^{\rm nor}_s = \frac{1}{\sqrt{\epsilon}}\left(5.25395-4.41951 \epsilon+O(\epsilon^2)\right)\,,\\
   &N=2: \quad  a^{\rm nor}_s = \frac{1}{204 \sqrt{14} \sqrt{\epsilon}}\left(4284-\epsilon \left(973+1530 \log (4)\right)+O(\epsilon^2)\right)~. 
    \end{split}
\end{equation}

\subsection{Free energy at order $\epsilon$}\label{SubSec:FreeEnergyInEpsExpansion}
We now compute the free energy of the normal universality class to order
$\epsilon$ at the fixed point, and organize the free energy as
\begin{gather}\label{FGNYep}
F=F_{\textit{tree}}+F_t+N F_{\Psi}+\sum \limits_{A=1}^{6}F^{(A)}_{\textit{2-loop}}+F_{\rm curv.}~.
\end{gather}
Here, the first four terms collect the classical saddle, the one-loop
determinants, and the two-loop corrections from the GNY interactions.
The final term, $F_{\rm curv.}$, denotes the contribution of the additional
curvature counterterms, which will be discussed separately in Section~ \ref{SubSec:CurvatureTerm}. 

The first three terms in \eqref{FGNYep} were considered in \cite{Giombi:2021cnr}:
\begin{align}
    &F_{\textit{tree}}=-\frac{3d^2(d-2)^2}{32g_{2,0}}{\rm V}_d\,,\label{Ftree}\\
   &F_t=F_D+\frac{{\rm V}_d}{2(4 \pi )^{d/2}}\int \limits_{\frac{d}{2}}^{\hat{\Delta}_t}d
\hat{\Delta}\left(2\hat{\Delta}-d+1\right)\frac{\Gamma\left(\hat{\Delta}\right)\Gamma\left(1-\frac{d}{2}\right)}{\Gamma\left(2-d+\hat{\Delta}\right)}\,,\label{Ft}\\
&  NF_{\Psi}=NF_{\rm free}-Nc_d\frac{\Gamma \left(1-\frac{d}{2}\right){\rm V}_d}{(4 \pi )^{d/2}}\int \limits_{0}^{\mu_0}dm \frac{   \Gamma \left(\frac{d}{2}+m \right)}{ \Gamma \left(1-\frac{d}{2}+m \right)}\,.
\label{NFPsi}
\end{align}
To order $\epsilon$, $F_D$ is given by \cite{Giombi:2025pxx}
\begin{align}\label{FDexp1}
F_D \stackrel{d=4-\epsilon}{=} \frac{1}{180\epsilon} +\frac{240\log(A)-480\zeta'(-3)-29-16\gamma_E}{2880} - \frac{\zeta(3)}{(4\pi)^2}- 0.003149\epsilon+\CO(\epsilon^2)\,~,
\end{align}
and $F_{\rm free}$ is the AdS free energy of a massless Dirac fermion, which can be obtained from the corresponding sphere free
energy. In our conventions, the latter is given by \cite{Giombi:2014xxa}
\begin{equation}
    F_{\Psi,S^d}
    =
    -\frac{c_d}{\sin\!\left(\frac{\pi d}{2}\right)\Gamma(1+d)}
    \int_0^1 du\, \cos\!\left(\frac{\pi u}{2}\right)\,
    \Gamma\!\left(\frac{1+d+u}{2}\right)
    \Gamma\!\left(\frac{1+d-u}{2}\right) .
    \label{FFermionOnSphere}
\end{equation}
To relate this result to EAdS$_d$, we use the fact that EAdS$_d$ is conformally equivalent to a hemisphere.
Imposing the conformal boundary condition on the hemisphere effectively removes
half of the modes of the Dirac operator on the sphere \cite{Sato:2021eqo}. Therefore, the free energy  of a massless Dirac spinor on EAdS$_d$ is  half of
the corresponding sphere free energy: 
\footnote{In Appendix~\ref{FreeEnergyMasslessFermions} we also compute
$F_{\rm free}$ directly to order $\epsilon$, without using the sphere
result.}
\begin{align}\label{Ffree}
F_{\rm free}=\frac{1}{2} F_{\Psi,S^d}=\frac{11}{180 \epsilon }-\frac{41}{432}+\frac{2 \log (A)}{3}-\frac{\zeta '(4)}{2 \pi ^4}-\frac{\gamma_E }{18}+\frac{\log (2 \pi )}{180} +0.04593 \epsilon+O(\epsilon^2)\,.
\end{align}

\begin{figure}
\centering
\begin{tikzpicture} [baseline={(0,0)}]
\draw[color=black,dashed] (0,0) circle [radius=0.5]; 
\draw[color=black,dashed] (1,0) circle [radius=0.5]; 
\draw (0.5,0) node[vertex]{} ;
\node at (0.5,-1) {$\CF_{\textit{2-loop}}^{(1)}$};
\end{tikzpicture} 
\qquad
\begin{tikzpicture} [baseline={(0,0)}]
\draw[color=black,dashed] (0,0) circle [radius=0.5]; 
\draw[color=black,dashed] (1.5,0) circle [radius=0.5]; 
\draw[dashed] (0.5,0) node[vertex]{} to (1,0) node[vertex]{};
\node at (0.75,-1) {$\CF_{\textit{2-loop}}^{(2)}$};
\end{tikzpicture} 
\qquad
\begin{tikzpicture} [baseline={(0,0)}]
\draw[dashed] (0,0) node[vertex]{} to (2,0) node[vertex]{} to [out=120, in=60] (0,0) to [out=-60, in=-120] (2,0);
\node at (1,-1) {$\CF_{\textit{2-loop}}^{(3)}$};
\end{tikzpicture} 
\\
\begin{tikzpicture} [baseline={(0,0)}]
\draw[color=black,dashed] (0,0) circle [radius=0.5]; 
\draw[ color=black] (1.5,0) circle [radius=0.5]; 
\draw[dashed] (0.5,0) node[vertex]{} to (1,0) node[vertex]{};
\node at (0.75,-1) {$\CF_{\textit{2-loop}}^{(4)}$};
\end{tikzpicture} 
\qquad
\begin{tikzpicture} [baseline={(0,0)}]
\draw[color=black] (0,0) circle [radius=0.5]; 
\draw[ color=black] (1.5,0) circle [radius=0.5]; 
\draw[dashed](0.5,0) node[vertex]{} to (1,0) node[vertex]{};
\node at (0.75,-1) {$\CF_{\textit{2-loop}}^{(5)}$};
\end{tikzpicture} 
\qquad
\begin{tikzpicture} [baseline={(0,0)}]
\draw[dashed](0,0) node[vertex]{} -- (2,0) node[vertex]{};
\draw[] (2,0) to[out=120, in=60] (0,0);
\draw[] (0,0) to[out=-60, in=-120] (2,0);
\node at (1,-1) {$\CF_{\textit{2-loop}}^{(6)}$};
\end{tikzpicture}
\caption{The two-loop contributions to the free energy. Dashed lines denote scalar propagators, while solid lines denote fermion propagators.}
\label{B1diagramsFreeEnergy}
\end{figure}
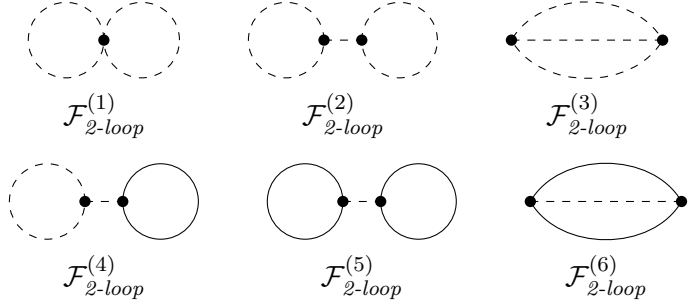

At the two-loop order, there are six diagrams depicted in Figure \ref{B1diagramsFreeEnergy}. Since the diagrams $\CF_{\textit{2-loop}}^{(1)}-\CF_{\textit{2-loop}}^{(6)}$ already start at order $\epsilon$, we may simply replace the bare couplings by the renormalized couplings. 
The first three were computed in \cite{Giombi:2025pxx}, and here we only provide the final result:
\begin{align}\label{CFold}
\CF_{\textit{2-loop}}^{(1)}+\CF_{\textit{2-loop}}^{(2)}+\CF_{\textit{2-loop}}^{(3)}= {\rm V}_d\left[ \frac{17g_2}{256 \pi ^4 \epsilon }+\left[f_{\textit{2-loop}}^{(1)}+f_{\textit{2-loop}}^{(2)}+f_{\textit{2-loop}}^{(3)}\right]_{\rm fin.}+O(\epsilon)\right]~,
\end{align}
where
\begin{gather}
    \left[f_{\textit{2-loop}}^{(1)}+f_{\textit{2-loop}}^{(2)}+f_{\textit{2-loop}}^{(3)}\right]_{\rm fin.}=-g_2\frac{(127-34 \Upsilon )}{512 \pi ^4}\,,
\end{gather}
with $\Upsilon$ given in \eqref{UpsilonDefinition}.
At large $N$ this contribution is of order $\frac{1}{N}$.
The remaining diagrams can be computed  using \eqref{Gpsi(x,x)} and \eqref{Gt(x,x)}
\begin{align}
    &\CF_{\textit{2-loop}}^{(4)}=\frac{g_{1}\sqrt{g_{2}} }{2}{\rm V}_d\sqrt{\frac{3d(d-2)}{2}} \frac{G_t(1)\langle \bar{\Psi}_0 \Psi_{0}\rangle}{\hat{\Delta}_t(\hat{\Delta}_t-d+1)}\,,\\
    &\CF_{\textit{2-loop}}^{(5)}=-\frac{g^2_{1}}{2}{\rm V}_d \frac{\langle \bar{\Psi}_0 \Psi_{0}\rangle^2}{\hat{\Delta}_t(\hat{\Delta}_t-d+1)}\,,\\
     &\CF_{\textit{2-loop}}^{(6)}=\frac{g^2_{1}N}{2}\int_{x,y} G_t(x,y)\tr{\left(G^+_{\mu}(x,y)G^+_{\mu}(y,x)\right)}\,,\label{F6diagram}
\end{align}
where $\mu=\frac{g_{1}}{\sqrt{g_{2}}}\sqrt{\frac{3d(d-2)}{2}}$. 

After expanding to order $\epsilon$ all contributions discussed above, the resulting expressions become rather lengthy. For this reason, we only outline the main steps of the computation and explain its structure, without presenting all intermediate expressions explicitly. 

First, in order to expand the tree-level contribution consistently to order $\epsilon$, we must use the two-loop renormalization of the bare couplings $g_{1,0}$ and $g_{2,0}$ \eqref{barecouplings}, and then perform the rescaling described in the beginning of the Section~\ref{Sec:Eps_Expansion}.
In this way, to the order relevant for our analysis, we obtain, introducing notation $G=\tfrac{g^2_1}{g_2}$:
\begin{align}
    -\frac{3d^2(d-2)^2}{32g_{2,0}}&=-3 g^2_1N^2\frac{ G(12 G-1)^2 }{8 \pi ^4 \epsilon^2}-9 g^2_1N\frac{(2 G-1) (12 G-1)}{32 \pi ^4 \epsilon^2}+\frac{9}{8 \pi ^2 \epsilon}-3N\frac{ G(6 G-1)}{\pi ^2 \epsilon}\nonumber\\
    &+9 g^2_1N^2\frac{ G(1-12 G)^2 }{16 \pi ^4 \epsilon}+3 g^2_1N\frac{(312 G^2-119G+6)}{64 \pi ^4\epsilon}-\frac{17 g_2 }{256 \pi ^4 \epsilon }+\left[f_{\rm tree}\right]_{\rm fin.}\,,
\end{align}
where $\left[f_{\rm tree}\right]_{\rm fin.}$ denotes the finite part,
\begin{align}
    \left[f_{\rm tree}\right]_{\rm fin.}&=-\frac{6}{g_{2}}+\frac{9 \epsilon}{g_{2}}-\frac{39 \epsilon^2}{8 g_{2}}+9 \left(1-\frac{13 \epsilon}{24}\right)\frac{ \left(N(48 G^2 -8 G)-3\right)}{16 \pi ^2}\nonumber\\
    &-3 g^2_1\frac{ \left(52G^2 N^2 (12 G-1)^2 +N\left(2088 G^3-462 G^2+3 G\right)-17\right)}{512 \pi ^4 G}\,.
\end{align}
When this expression is evaluated at the fixed point, this is in fact the only place where the order $\epsilon^3$ correction to $g_{2,\star}$ is needed, due to the presence of the term $-\tfrac{6}{g_2}$.

To the order of interest, $F_t$ was computed recently in \cite{Giombi:2025pxx} and is given by
\begin{gather}
    F_t
=F_D+{\rm V}_d\left(-\frac{9}{8\pi^2\epsilon}+\left[f_t\right]_{\rm fin.}\right)+O(\epsilon^2)\,,
\end{gather}
where
\begin{gather}
    \left[f_t\right]_{\rm fin.}=\frac{3 \left(24 \log (A)-\frac{2 \zeta (3)}{\pi ^2}+6 \gamma_E +31-22 \log (2)-16 \log (\pi )\right)}{32 \pi ^2}-0.0677065 \epsilon\,.
\end{gather}

Next, we turn to the one-loop fermionic contribution to the free energy. One technical complication is that the mass $\mu_0$ depends on a ratio of couplings. We therefore begin by expanding the integrand in \eqref{NFPsi} to order $\epsilon$:
\begin{gather}
Nc_d\frac{\Gamma \left(1-\frac{d}{2}\right)}{(4 \pi )^{d/2}}\frac{\Gamma \left(\frac{d}{2}+m \right)}{\Gamma \left(1-\frac{d}{2}+m \right)}
=\frac{W_{-1}(m)}{\epsilon}+W_{0}(m)+\epsilon W_{1}(m)+O(\epsilon^2)\,.
\end{gather}
The terms $W_{-1}$ and $W_0$ can be integrated analytically and then
expanded in the renormalized couplings using the prescription described
above to the required order. The contribution of $W_1$ is finite. We separate it into an
analytic part and a remaining one-dimensional integral, which we evaluate
numerically at fixed $N$. With $G=g_1^2/g_2$, this gives
\begin{align}
  NF_{\Psi}&=NF_{\rm free}+{\rm V}_d\Bigg(N\frac{3G (6 G-1) }{\pi ^2 \epsilon}+3 g^2_1N^2\frac{ G (12 G-1)^2 }{4 \pi ^4 \epsilon^2}+9 g^2_1N\frac{ (2 G-1) (12 G-1)}{16 \pi ^4 \epsilon^2}\nonumber\\
    &-\frac{g^2_1\sqrt{G}N}{16 \pi ^4 \epsilon }\left(144 G^{3/2}+36 \sqrt{3} G-3 \sqrt{G}-\sqrt{3}\right)\left((12 G-1) N+\frac{3 (2 G-1)}{4 G}\right)\nonumber\\
    &-\frac{3g^2_{1}N}{32 \pi ^4 \epsilon}\left(4 NG(12 G-1)^2+3(2 G (12 G-7)+1)\right)\left(2 H_{\sqrt{12G}-2}-\Upsilon\right)+\left[f_{\Psi}\right]_{\rm fin.}\Bigg)\,,
\end{align}
where the finite part is written as
\begin{gather}
    \left[f_{\Psi}\right]_{\rm fin.}=\left[f_{\Psi}\right]_{\rm an.}+\epsilon N \int_0^{\frac{6}{S_N}} \frac{y \left(y^2-1\right) \psi ^{(0)}(y+1)^2}{4 \pi ^2} \, dy\,,
\end{gather}
where $S_N$ is given in \eqref{SNDefinition}. The explicit expression for the analytic part
$\left[f_{\Psi}\right]_{\rm an.}$ is rather lengthy, so we do not display
it here.

Next, the contributions $\CF_{\textit{2-loop}}^{(4)}$ and $\CF_{\textit{2-loop}}^{(5)}$ are straightforward to evaluate with results (using that $G=\tfrac{g^2_{1}}{g_2}$):
\begin{align}
    \CF_{\textit{2-loop}}^{(4)}&={\rm V}_d\Bigg(g^2_1N\frac{9  (12 G-1)}{16 \pi ^4 \epsilon^2}-\frac{3 g^2_1N \left(276 G^{3/2}+36 \sqrt{3} G-14 \sqrt{G}-\sqrt{3}\right)}{64 \pi ^4 \epsilon\sqrt{G}}\nonumber\\
    &
    -\frac{9 g^2_1N (12 G-1) }{16 \pi ^4 \epsilon}\left(H_{ \sqrt{12G}-2}-\Upsilon\right)+\left[f_{\textit{2-loop}}^{(4)}\right]_{\rm fin.}\Bigg)\,,
\end{align}
where $\Upsilon$ in given in \eqref{UpsilonDefinition} and
\begin{align}
    \left[f_{\textit{2-loop}}^{(4)}\right]_{\rm fin.}&=g^2_{1}N
    \frac{ \left(144 \left(80+23 \pi ^2\right) G+6660 \sqrt{3} \sqrt{G}-\frac{95 \sqrt{3}}{\sqrt{G}}-276 \pi ^2+1560\right) }{2560 \pi ^4}\nonumber\\
    &+3 g^2_1N \frac{ \left(276 G^{3/2}+36 \sqrt{3} G-14 \sqrt{G}-\sqrt{3}\right) }{64 \pi ^4 \sqrt{G}}\left(H_{ \sqrt{12G}-2}-\Upsilon\right)\\
    &+9 g^2_1N\frac{ (12 G-1) }{32 \pi ^4}\left(H_{\sqrt{12G}-2}-\Upsilon \right){}^2+27 \sqrt{3}  g^2_{1}N\frac{\sqrt{G} (12 G-1) }{64 \pi ^4}\psi ^{(1)}\left(\sqrt{12G}-1\right)\nonumber\,,
\end{align}
as well as
\begin{align}
    \CF_{\textit{2-loop}}^{(5)}&={\rm V}_d\Bigg(-3g^2_{1}N^2\frac{ G(12 G-1)^2 }{8 \pi ^4 \epsilon^2}+g^2_{1}N^2\frac{ (12 G-1) \left(36 \sqrt{3} G^{\frac{3}{2}}+36 G^2+6 G-\sqrt{3} \sqrt{G}\right)}{16 \pi ^4 \epsilon}\nonumber\\
    &+3 g^2_1N^2\frac{ G(12 G-1)^2  }{8 \pi ^4 \epsilon}\left(2H_{\sqrt{12 G}-2}-\Upsilon\right)+\left[f_{\textit{2-loop}}^{(5)}\right]_{\rm fin.}\Bigg)\,,
\end{align}
where
\begin{align}
    \left[f_{\textit{2-loop}}^{(5)}\right]_{\rm fin.}&=-g^2_{1}N^2\frac{\left(2 \pi ^2 G (12 G-1)^2+36 G (4 G (3 G+14)-3)+1\right)}{128 \pi ^4}\nonumber\\
    &+\sqrt{3} g^2_1N^2\frac{ \sqrt{G} (11-48 G (9 G+5))}{128 \pi ^4}-3 g^2_1N^2\frac{ G(12 G-1)^2  }{16 \pi ^4}\left(2H_{\sqrt{12 G}-2}-\Upsilon\right)^2\nonumber\\
    &-g^2_{1}N^2\frac{ \sqrt{G} (12 G-1) \left(36 G^{3/2}+36 \sqrt{3} G+6\sqrt{G}-\sqrt{3}\right)}{16 \pi ^4}\left(2H_{\sqrt{12 G}-2}-\Upsilon \right)\\
    &-9 \sqrt{3} g_1^2N^2\frac{  G^{\frac{3}{2}}(12 G-1)^2  }{16 \pi ^4}\psi ^{(1)}\left( \sqrt{12G}-1\right)\nonumber\,.
\end{align}Finally, $\CF_{\textit{2-loop}}^{(6)}$ requires a more careful analysis, and we discuss it in Appendix \ref{App:SubtractionFunction} with result 
\begin{align}\label{F6diagramFinal}
    \CF_{\textit{2-loop}}^{(6)}&={\rm V}_d\Bigg(-9 g^2_1N\frac{ (2 G+1) (12 G-1)}{32 \pi ^4 \epsilon^2}+3 g^2_1N\frac{ \left(72 \sqrt{3} G^{\frac{3}{2}}-24 G^2+245 G-\sqrt{12G}-17\right)}{64 \pi ^4 \epsilon}\nonumber\\
    &+9 g^2_1N\frac{G (12 G-1)  )}{16 \pi ^4 \epsilon}\left(2 H_{ \sqrt{12G}-2}-\Upsilon\left(1+\frac{1}{2G}\right)\right)+\left[f_{\textit{2-loop}}^{(6)}\right]_{\rm fin.}\Bigg)\,.
\end{align}
We do not present the explicit form of
$\left[f_{\textit{2-loop}}^{(6)}\right]_{\rm fin.}$, since it is rather
lengthy. This finite part contains both an analytic contribution and a
finite one-dimensional integral. For generic $N$, the latter
cannot be evaluated analytically, so we compute it numerically for each
fixed value of $N$. In the large $N$
expansion, the leading contribution of order $N^0$ simplifies and can be obtained analytically.

Putting everything together, we find that all divergences cancel, providing a nontrivial consistency check. The free energy is therefore given by
\begin{gather}
    F=F_D+NF_{\rm free}+{\rm V}_d\left(\left[f_{\rm tree}\right]_{\rm fin.}+\left[f_{t}\right]_{\rm fin.}+\left[f_{\Psi}\right]_{\rm fin.}+\sum \limits_{A=1}^{6}\left[f_{\textit{2-loop}}^{(A)}\right]_{\rm fin.}\right)+F_{\rm curv.}.
    \label{FfullEpsExpansion}
\end{gather}
Before discussing the contribution of the curvature terms to the free
energy, it is useful to compare the result obtained so far with the
large $N$ expansion. Combining the leading order result \cite{Giombi:2021cnr}, which to order $\epsilon$ takes the form
\begin{gather*}
    F^{(0)}_{\rm nor}=NF_{free}
    +N{\rm V}_d\left(-\frac{1}{8 \pi ^2 \epsilon}+\frac{17-3 \log \left(4 \pi  e^{-\gamma_E }\right)-72 \log (A)+\frac{18 \zeta (3)}{\pi ^2} }{48 \pi ^2}-0.00470866 \epsilon\right)\,,
    \label{FB1LeadingOrder}
\end{gather*}
with subleading result \eqref{FB1order1}, we then find after taking large $N$ limit of \eqref{FfullEpsExpansion}
\begin{gather}
    F^{(0)}_{\rm nor}+F_{\rm nor}^{(1)}-F=\frac{\epsilon}{16}-F_{\rm curv.}+O\left(\epsilon^2,\frac{1}{N}\right).
    \label{LargeNepsExpansComparisonFreeEnergy}
\end{gather}
This comparison shows that the contribution evaluated so far differs
from the large $N$ result by the finite term $\tfrac{\epsilon}{16}$. As we show in
the next subsection, this difference is exactly accounted for by the
curvature counterterm contribution.

\subsubsection{The curvature counterterm}\label{SubSec:CurvatureTerm}
In order to renormalize a quantum field theory on a curved manifold, one has to include all possible curvature counterterms. In our case, with the manifold being  maximally symmetric, there are two curvature counterterms that are marginal in $4d$, which take the form $s_0^2\mathcal R$ and $\mathcal R^2$ respectively, where $\mathcal R$ is the Ricci scalar of EAdS$_d$. According to \cite{Jack:1990eb,Fei:2016sgs}, the pure curvature term $\mathcal R^2$ does not contribute to the free energy at order $\epsilon$. Therefore, we will drop this term, and consider only $s_0^2\mathcal R$ in \eqref{GNYAdS}
\begin{gather}
    S_{\rm curv.}=\frac{1}{2}\eta_0\int d^dx\sqrt{g}\,s_0^2 H\,, \qquad H=\frac{\mathcal{R}}{d-1}\,.
    \label{ActionCurvatureTerm}
\end{gather}
The renormalization of the curvature coupling takes the form \cite{Jack:1990eb,Fei:2015oha}:
\begin{gather}
    \eta_0=\frac{\eta+L_{\eta}}{Z_{s^2}}\,, \qquad \quad L_{\eta}=\sum\limits_{k=1}^{+\infty} \epsilon^{-k}L^{(k)}_{\eta}(g^2_1,g_2)\,, \quad Z_{s^2}=1-\frac{N g^2_1 +g_2}{(4\pi)^2 \epsilon}\,,
    \label{RenormalizationEta0}
\end{gather}
where in the leading order in coupling constants we have
\begin{gather}
    L^{(1)}_{\eta}(g^2_1,g_2)=\frac{1}{3(16\pi^2)^3}\left(6N^2 g_1^6-\frac{1}{2}g_1^2 g_2^2N\right)\,.
\end{gather}
The $\beta$ function of the corresponding renormalized coupling $\eta$ was studied in \cite{Jack:1990eb} and takes the form\footnote{To determine $\hat{\beta}_\eta(g_1,g_2)$, we use eq.~(7.22) of \cite{Jack:1990eb}. The relation between the quantity $\beta^\eta(\phi)$ appearing there and our notation follows from the footnote below eq.~(6.29) of \cite{Jack:1990eb}, from which we infer
$2\beta^\eta(\phi)=\hat{\beta}_\eta(g_1,g_2)\phi^2$.}
\begin{gather}
\beta_\eta=\gamma_{s^2}\eta+\hat{\beta}_\eta(g_1,g_2)\,, 
\end{gather}
and $\hat{\beta}_\eta(g_1,g_2)$ is fixed by $L^{(1)}_{\eta}(g^2_1,g_2)$
\begin{gather}
\hat{\beta}_\eta(g_1,g_2)=\left(g^2_{1}\frac{\partial}{\partial g^2_{1}}+g_{2}\frac{\partial}{\partial g_{2}}\right)L^{(1)}_{\eta}(g^2_1,g_2)=\frac{1}{\left(16 \pi ^2\right)^3}\left(6 g_1^6 N^2-\frac{1}{2} g_1^2 g_2^2 N\right)\,.
\end{gather}
The anomalous dimension $\gamma_{s^2}$ is given by \cite{Fei:2016sgs}
\begin{align}
    \gamma_{s^2}=\frac{\sqrt{4 N^2+132 N+9}+10 N+3}{6 (2 N+3)}\epsilon+\mathcal{O}(\epsilon^2)\,.
\end{align}
Evaluating these expressions at the fixed point, we find
\begin{gather}
    \eta_\star=-\frac{\hat{\beta}_\eta(g_{1,\star},g_{2,\star})}{\gamma_{s^2}}
    =\frac{ N \left(2N+15-\sqrt{4 N^2+132 N+9}\right)}{48 (2N+3)^2}\epsilon^2\,.
\end{gather}

Since $\eta_\star\sim \epsilon^2$, in principle it would be sufficient to keep only the leading contribution of $\langle s_0^2\rangle$ to the free energy, namely, replacing  it with $s^2_{\star,0}\sim \tfrac{1}{g_{2,\star}}\sim \frac{1}{\epsilon}$. However, in order to explicitly verify the cancellation of UV divergences in subleading contributions coming from one point function $\langle t \rangle$ as well as $\langle t^2 \rangle$, we consider the following contributions to the free energy, using \eqref{s_star} together with \eqref{1pftAsSumOfT1PT2}:
\begin{gather}
    F_{\rm curv.}=\frac{1}{2}\eta_0{\rm V}_d H\left[s_{\star,0}^2+2s_{\star,0}\left(\mathcal{T}_1+\mathcal{T}_2\right)+G_{t}(1)+\mathcal{O}(g^2_1,g_2)\right]\,.
\end{gather}
After expressing everything in terms of renormalized couplings, all UV divergences must cancel, since there are no additional counterterms at this order that could compensate them. Using the renormalization discussed above, one indeed finds finite answer to the leading order\footnote{This result could in fact be obtained directly from \eqref{ActionCurvatureTerm} to the order relevant for our analysis. Indeed, one may first neglect $L_\eta$ in \eqref{RenormalizationEta0}, since it will be cancelled by higher order terms. Then, using the relation between the bare and renormalized composite operators, $s_0^2=Z_{s^2}s^2$, one finds that the curvature term contributes to the free energy as $\frac{1}{2}\eta {\rm V}_d H\,\langle s^2\rangle$. Furthermore, since we are only interested in the leading contribution, one may replace $\langle s^2\rangle$ by $\langle s\rangle^2$. This is because the one-point function $\langle s\rangle$ gives the leading contribution to $\langle s^2\rangle$, of order $\tfrac{1}{g_{2,\star}}$, thereby reproducing \eqref{DeltaFCurvatureTerm}. }
\begin{gather}
     F_{\rm curv.}=\frac{1}{2}\eta_\star{\rm V}_d H\left[\frac{12}{g_{2,\star}}+\mathcal{O}\left(1\right)\right]=\frac{ \left(6N+3-\sqrt{4 N^2+132 N+9}\right)}{32(2 N+3)}\epsilon+\mathcal{O}(\epsilon^2)\,,
    \label{DeltaFCurvatureTerm}
\end{gather}
which has the following
large $N$ expansion
\begin{gather}
    F_{\rm curv.}
    =
    \frac{\epsilon}{16}
    +\mathcal{O}\left(\frac{1}{N}\right).
\end{gather}
Together with \eqref{LargeNepsExpansComparisonFreeEnergy}, this shows
that the full free energy reproduces the large $N$ expansion to the order
considered.

Combining \eqref{FfullEpsExpansion} with the curvature contribution
\eqref{DeltaFCurvatureTerm}, we arrive at the final result for the free
energy through order $\epsilon$
\begin{gather}
    F=\frac{1}{\epsilon}
    \frac{(11 N+1) S_N^2-540 (2 N+3)}{180 S_N^2}
    +F_{\epsilon^0}+\epsilon F_{\epsilon}\,,
\end{gather}
where $S_N$ is defined in \eqref{SNDefinition}. The coefficient of
the $1/\epsilon$ term takes the compact form. This term receives
contributions only from the Weyl anomalies of the conformally coupled
scalar and the massless Dirac fermions, together with the tree-level
contribution. The finite term $F_{\epsilon^0}$ is an explicit analytic
function of $N$, whereas $F_{\epsilon}$ contains both analytic terms and
finite one-dimensional integrals, which we evaluate numerically for each
fixed value of $N$. We do not provide final expressions for $F_{\epsilon^0}$ and $F_{\epsilon}$ as they are lengthy. The numerical values for several small values
of $N$ are collected in Table~\ref{tab:f_expansion}.

\begin{table}[t]
\centering
\renewcommand{\arraystretch}{1.35}
\setlength{\tabcolsep}{14pt}
\begin{tabular}{|c|c|}
\hline
$N$ & $F$ \\
\hline
$\tfrac{1}{4}$ & $\displaystyle -1.14583\epsilon^{-1}+1.43007 -3.78925\epsilon$ \\
\hline
$1$ & $\displaystyle -1.0835\epsilon^{-1}+1.45508 -4.2527\epsilon$ \\
\hline
$2$ & $\displaystyle -1.18472\epsilon^{-1}+1.54475-4.75653\epsilon$ \\
\hline
$3$ & $\displaystyle -1.31111\epsilon^{-1}+1.59834-5.09622\epsilon$ \\
\hline
$4$ & $\displaystyle -1.44098\epsilon^{-1}+1.62992-5.36331\epsilon$ \\
\hline
$5$ & $\displaystyle -1.57014\epsilon ^{-1}+1.64749-5.59492\epsilon  $ \\
\hline
\end{tabular}
\hspace{0.1cm}
\begin{tabular}{|c|c|}
\hline
$N$ & $F $ \\
\hline
$6$ & $\displaystyle -1.69761\epsilon^{-1}+1.65541-5.80754\epsilon$ \\
\hline
$7$ & $\displaystyle -
1.82324\epsilon^{-1}+1.65626-6.0093\epsilon$ \\
\hline
$8$ & $\displaystyle -1.94713\epsilon^{-1}+1.65172-6.20453 \epsilon$ \\
\hline
$9$ & $\displaystyle -2.06944\epsilon^{-1}+1.64291-6.39571\epsilon$ \\
\hline
$10$ & $\displaystyle -2.19035\epsilon^{-1}+1.63064-6.58431\epsilon$ \\
\hline
$11$ & $\displaystyle -2.31001\epsilon^{-1}+1.61553-6.77124\epsilon$ \\
\hline
\end{tabular}

\caption{The free energy \eqref{FfullEpsExpansion} for some small $N$.}
\label{tab:f_expansion}
\end{table}

\subsubsection{The boundary central charge}
Given a $3d$ boundary  CFT, the boundary central charge $c_{\rm bdry}$ is a boundary RG monotone, decreasing from UV to IR  \cite{Jensen:2015swa, Herzog:2017kkj}. By placing a CFT in a $3d$ hyperbolic space, the coefficient of the $\log(R)$ term in the free energy also captures the boundary central charge, where $R$ is an IR cut-off. A dimensional continuation of $c_{\rm bdry}$ was proposed by 
\cite{Kobayashi:2018lil}:
\begin{align}\label{ts}
\tilde{s}
= - \sin\!\left(\frac{\pi (d-1)}{2}\right)
\left(F-\frac{1}{2}F_{S^d}\right)\,,
\end{align}
where $F$ denotes the free energy of the boundary CFT in the $d$ dimensional hyperbolic space, and $F_{S^d}$ is the free energy of the bulk CFT in a unit $d$ dimensional sphere. In particular, in $3d$, $\tilde s$ is proportional to the boundary central charge, i.e. $\tilde s\stackrel{d=3}{=} \frac{\pi}{6} c_{\rm bdry}$, and in $2d$, it gives the logarithm of the boundary $g$-function, i.e. $\tilde s\stackrel{d=2}{=} \log(g)$. In a recent work \cite{Giombi:2025pxx}, this quantity has been used to estimate boundary central charges of the 3$d$ critical O$(N)$ model. See \cite{Feng:2026iii} for numerical comparison with fuzzy sphere regularization. 

We have computed $F$ for the normal boundary condition of the GNY model to order $\epsilon$.
The sphere free energy of the GNY CFT is known to order $\epsilon^2$ \cite{Fei:2016sgs}. Here, we present $F_{S^d}$ to order $\epsilon$: 
\begin{align}\label{FS}
F_{S^d} = N F_{\Psi,S^d} + F_{\rm conf.} - \frac{N}{24(2N+3)}\,\epsilon + O(\epsilon^2)\, .
\end{align}
where $F_{\Psi,S^d}=2F_{\rm free}$ is the sphere free energy of a massless Dirac spinor, and $F_{\rm conf.}$ is the sphere free energy of  a conformally coupled scalar \cite{Giombi:2014xxa}
\begin{align}
F_{\rm conf.}
=\frac{1}{90 \epsilon }+0.013114+0.013255 \epsilon~. \label{FScalarOnSphere}
\end{align}
Plugging \eqref{FfullEpsExpansion} and \eqref{FS} into \eqref{ts}, we obtain the $\epsilon$ expansion of $\tilde s$ for the normal boundary condition of the GNY CFT. For example, at $N=\frac{1}{4}$, the result reads
\begin{gather}\label{tsSUSY}
    \tilde{s}^{\rm nor}_{N=1/4}=-\frac{7}{6\epsilon}+1.41115\, -2.36656 \epsilon~.
\end{gather}
The numerical values of $ \tilde{s}^{\rm nor}_{N}$ at other small $N$ are reported in Table \ref{tab:s_expansion}.
Because the Weyl anomalies cancel in the combination $F-\frac{1}{2}F_{S^d}$, 
the $1/\epsilon$ term is solely due to the tree-level action.

\begin{table}[t]
\centering
\renewcommand{\arraystretch}{1.35}
\setlength{\tabcolsep}{14pt}
\begin{tabular}{|c|c|}
\hline
$N$ & $\tilde s$ \\
\hline
$\tfrac{1}{4}$ & $\displaystyle -1.1667\epsilon^{-1}+1.41115\, -2.36656 \epsilon$ \\
\hline
$1$ & $\displaystyle -1.15017\epsilon^{-1}+1.3991-2.88213 \epsilon$ \\
\hline
$2$ & $\displaystyle -1.3125\epsilon^{-1}+1.43934-3.22984 \epsilon$ \\
\hline
$3$ & $\displaystyle -1.5\epsilon^{-1}+1.44351-3.38315 \epsilon$ \\
\hline
$4$ & $\displaystyle -1.69098\epsilon^{-1}+1.42566-3.45993 \epsilon$ \\
\hline
$5$ & $\displaystyle-1.88125\epsilon^{-1}+1.3938-3.5023 \epsilon  $ \\
\hline
\end{tabular}
\hspace{0.1cm}
\begin{tabular}{|c|c|}
\hline
$N$ & $\tilde s $ \\
\hline
$6$ & $\displaystyle-2.06984\epsilon^{-1}+ 1.3523-3.52788 \epsilon$ \\
\hline
$7$ & $\displaystyle-2.25658\epsilon^{-1}+ 1.30373-3.54494\epsilon$ \\
\hline
$8$ & $\displaystyle -2.44158\epsilon^{-1}+1.24976-3.55768 \epsilon$ \\
\hline
$9$ & $\displaystyle -2.625\epsilon^{-1}+1.19152-3.56834\epsilon$ \\
\hline
$10$ & $\displaystyle -2.80702\epsilon^{-1}+1.12983-3.57819\epsilon$ \\
\hline
$11$ & $\displaystyle -2.98779\epsilon^{-1}+1.06529-3.58793 \epsilon$ \\
\hline
\end{tabular}

\caption{The numerical values of $\tilde s_{N}^{\rm nor}$ \eqref{ts} for some small $N$.}
\label{tab:s_expansion}
\end{table}

\subsection{Pad\'e resummation}
Formally setting $N=\frac{1}{4}$, the IR fixed point of the $2d$ GNY model is believed to be described by the tricritical Ising minimal model $M(4, 5)$ \cite{Grover:2013rc,Shimada:2015gda,Fei:2016sgs}, which admits 6 boundary Cardy states. The properties of these states are discussed in detail in \cite{Nepomechie:2001bu,Balaska:2006yn,Iino:2019ogl}.
We identify the normal boundary condition of the GNY model with the most stable Cardy states of the $M(4,5)$ minimal model. Following the notations of \cite{Balaska:2006yn}, we denote such states by $|I\rangle$ and $|\varepsilon''\rangle$. They are related by the $\mathbb Z_2$ symmetry of $M(4,5)$, which is invisible in the GNY lagrangian description. The $g$-function for both states is
\begin{align}\label{gfun}
g_{|I\rangle} =g_{|\varepsilon''\rangle}= 5^{-\frac{1}{4}}\sqrt{\sin\left(\frac{4\pi}{5}\right)}\approx 0.5127~.     
\end{align}
In addition, according to the identification of \cite{Fei:2016sgs}, the canonically  normalized field $s$ corresponds to the Virasoro primary operator $\varepsilon$ in $M(4, 5)$, with dimension $(h,\bar h) = (\frac{1}{10},\frac{1}{10})$. The one-point function of $\varepsilon$ in the half-space with either $|I\rangle$ or $|\varepsilon''\rangle$ boundary condition is \cite{Balaska:2006yn}
\begin{align}\label{onept}
    \langle \varepsilon(z)\rangle_{I} =  \langle \varepsilon(z)\rangle_{\varepsilon''} = \frac{a_{\varepsilon}}{(2 y)^{\frac{1}{5}}}\,, \quad a_{\varepsilon} = \sqrt{\frac{\sin(\frac{2\pi}{5})}{\sin(\frac{4\pi}{5})}}\approx 1.27\,.
\end{align}
Applying the [0,1] Pad\'e resummation to $\sqrt{\epsilon}a^{\rm nor}_s$, with $a^{\rm nor}_s$ given by \eqref{assusy}, leads to 
\begin{align}
    \text{Pad\'e}_{[0, 1]}: \quad a^{\rm nor}_s(d) = \frac{2\sqrt{7}}{(1+\frac{27}{28}\epsilon)\sqrt{\epsilon}}~.
\end{align}
At $d=2$, this Pad\'e approximant gives $a^{\rm nor}_s(2)\approx 1.28$, which deviates from the exact value $a_\varepsilon$ by less than 1\% deviation. We can also impose the exact value in $2d$ as a constraint, which allows us to apply two-sided $[1,1]$ or $[0,2]$ Pad\'e approximant to the $\epsilon$-expansion. The two Pad\'e resummations appear to make approximately identical predictions for the one-point function  in 3$d$. The Pad\'e estimate of the  one-point coefficient of the $s$ operator in the normal boundary phase of the $3d$ super-Ising universality class is
\begin{align}
    3d\,\,\text{Super-Ising}: \qquad a^{\rm nor}_s \approx 2.69~. 
\end{align}
We also give the Pad\'e estimate of $a^{\rm nor}_s$ in $3d$ for some higher $N$  in the following table
\begin{center}
  \begin{tabular}{|c|c|c|c|c|}
\hline
     $N$& $\frac{1}{4}$ & $\frac{1}{2}$&   $1$ & 2  \\
     \hline
     $a^{\rm nor}_s(d=3)$ & 2.69 & 2.71 & 2.85& 3.26\\
     \hline
\end{tabular}  
\end{center}

Next, we consider the Pad\'e resummation of the $\tilde{s}$ function. We apply the Pad\'e approximant to $\epsilon\tilde{s}$ and then divide the resummed result by $\epsilon$. For the $\epsilon$-expansion \eqref{tsSUSY}, corresponding to $N=\frac{1}{4}$, the $[1,1]$ Pad\'e approximant yields $-0.26$ and the $[0,2]$ Pad\'e approximant yields $-0.5$ in $2d$. The exact $2d$ value, on the other hand, is $\log g_{|I\rangle}\approx -0.67$, where we have used \eqref{gfun}. The $[0,2]$ Pad\'e approximant is therefore significantly more accurate than the $[1,1]$. Based on the $[0,2]$ Pad\'e approximant, we obtain the estimate $-1.35$ for the boundary central charge in $3d$. Alternatively, we can use the exact $2d$ value as a constraint and apply the two-sided Pad\'e to $\epsilon\tilde{s}$. Among the three types of two-sided Pad\'e, i.e.\ $[m,3-m]$ with $m=0,1,2$, the $[0,3]$ and $[1,2]$ approximants are extremely close to each other and give
\begin{align}\label{bcSUSY}
  3d\,\,\text{Super-Ising}:\qquad  c^{\rm nor}_{\rm bdry} \approx -1.38
\end{align}
for the normal boundary condition of the $3d$ super-Ising universality class. This value is also close to the estimate based on the one-sided $[0,2]$ Pad\'e. The two-sided $[2,1]$ Pad\'e yields a value far from \eqref{bcSUSY}, and we therefore consider it an unreliable estimate.

Let's also comment on the Pad\'e resummation of the $\tilde s$ function for the ordinary phase at $N=\frac{1}{4}$. Combining \eqref{Fordexp} and \eqref{FS}, we obtain 
\begin{align}
    \tilde s^{\rm ord}_N = -0.007612 - 0.009776 \epsilon+\frac{\left(96 N^2+\sqrt{4 N^2+132 N+9}+46 N+3\right) \epsilon }{576 (2 N+3)}+O(\epsilon^2)~.
\end{align}
At $N=1/4$, it reduces to $  \tilde s^{\rm ord}_{N=1/4} = -0.007612 + 0.003616 \epsilon+O(\epsilon^2)$. In $2d$, we expect the ordinary phase to correspond to the Cardy states $|\varepsilon\rangle$ and $|\varepsilon'\rangle$ of the tricritical Ising model \cite{Balaska:2006yn}, which contain one relevant direction of dimension $3/5$. They have the same $g$ function 
\begin{align}
    g_{|\varepsilon\rangle} = g_{|\varepsilon'\rangle} = \frac{1}{5^{1/4}}\frac{\sin(\frac{2\pi}{5})}{\sqrt{\sin(\frac{4\pi}{5})}}\approx 0.83~.
\end{align}
Imposing this $2d$ constraint, we can apply two-sided Pad\'e resummation to $\tilde s^{\rm ord}_{N=1/4}$. The [1,1] Pad\'e approximant has a pole for $2<d<4$, so it is unreliable. The [0,2] Pad\'e approximant gives the following estimate of the boundary central charge:
\begin{align}
  3d\,\,\text{Super-Ising}:\qquad  c^{\rm ord}_{\rm bdry} \approx -0.015~.
\end{align}

\section*{Acknowledgements}

We thank Yifan Wang and Fedor Popov for discussions. Z.S.\ is supported by the U.S.\ Department of Energy grant DE-SC0009988 and the Sivian Fund.

\appendix

\section{$\beta$ functions of the GNY model}\label{betareview}
In this appendix, we summarize some useful results regarding the renormalization of the GNY model in $d=4-\epsilon$ dimensions.

The $\beta$-functions of the GNY model are known to the five-loop order \cite{Gracey:2025aoj}. For the purpose of this paper, it suffices to know the three-loop $\beta$-functions. We present them in terms of the rescaled couplings $y=\frac{g_1^2}{8\pi^2}$ and $ \lambda=\frac{g_2}{192\pi^2}$:
\begin{align}
    \beta_{y}=&-\epsilon y+(2 N+3) y^2+24 \lambda ^2 y-24 \lambda  y^2-\left(6 N+\frac{9}{8}\right) y^3-216 \lambda ^3 y-3 (30 N-91)  \lambda ^2y^2\nonumber\\ 
    &+18  (5 N+7) \lambda y^3 +\frac{2N (112N+432 \zeta(3)+67)+912 \zeta(3)-697}{64} y^4\,,
\end{align}
and
\begin{align}
    \beta_{\lambda}=&-\epsilon\lambda+36\lambda^2+4N\lambda y-Ny^2-816\lambda^3-72N\lambda^2y+7N\lambda y^2+4Ny^3\nonumber\\
   &+216 \lambda ^4 (96\zeta(3)+145)+1548N\lambda^3y-\frac{3N(72 N-648\zeta(3)-361)}{2}\lambda ^2y^2 \\
   &+\frac{N(1736N-1872\zeta(3)-4395)}{16}  \lambda  y^3-\frac{N(628N+384 \zeta(3)-5)}{32} y^4\,\nonumber.
\end{align}

We also give the precise relations between the bare couplings and renormalized couplings up to the two-loop order
\begin{gather}
 g^2_{1,0}=\mu^\epsilon \left(g^2_1+\frac{\left(N+6\right) g^4_1}{(4\pi)^2\epsilon}+\frac{1}{2(4 \pi )^4 \epsilon }\left(-\frac{3}{2}(4 N+3)g_1^6  -4g_1^4 g_2 +\frac{1}{6}g^2_1 g_2 ^2 \right)+\frac{(N+6)^2 g_1^6 }{(4 \pi )^4 \epsilon^2 }\right)\,, \nonumber \\
    g_{2,0}=\mu^{\epsilon}\Bigg(g_2+\frac{3g^2_2+2N g_2 g^2_1-12 N g^4_1}{(4\pi)^2\epsilon}+\frac{1}{2(4\pi)^4 \epsilon}\left(96 N g_1^6  +7Ng_1^4 g_2 -3Ng^2_1 g_2 ^2 - \frac{17g^3_2}{3}\right)\nonumber\\
    +\frac{1}{(4\pi)^4 \epsilon^2}\left(-24 N(N+3) g_1^6  +3N(N-10)g_1^4 g_2 +9Ng^2_1 g_2 ^2  +9g^3_2\right)\Bigg)\,.
    \label{barecouplings}
\end{gather}
where $\mu$ is an arbitrary energy scale. In Section \ref{SubSec:1pfInEpsExpansion}, it is enough to consider one-loop renormalization, while in Section \ref{SubSec:FreeEnergyInEpsExpansion} we will need to use the two-loop results for correctly taking into account all the divergences.

The explicit expressions of the nontrivial fixed-points to order $\epsilon^3$ can be found in  \cite{PhysRevB.96.165133}. Here we give the fixed point at the leading order
\begin{align}\label{fp}
   \frac{(g_{1\star})^2}{(4\pi)^2}=\frac{\epsilon}{4 N+6}, \quad \frac{g_{2\star}}{(4\pi)^2}=\frac{\left(\sqrt{4 N^2+132 N+9}-2 N+3\right) \epsilon }{6 (2 N+3)}~. 
\end{align}
Note that for large $N$ and small $\epsilon$, $g_{1\star}\sim \sqrt{\epsilon/N}$ and $g_{2\star} \sim \epsilon/N$.

\section{Harmonic analysis in AdS}\label{app:harmonic}

On $S^2$, the SO$(3)$ invariant two-point functions can be expanded into Legendre polynomials. The AdS$_d$ counterparts of Legendre polynomials are the so-called harmonic functions $\Omega_\nu(X, Y), \nu\in\mathbb R$, whose explicit form is given by \cite{Costa_2014}
\begin{align}\label{Omeganu}
\Omega_\nu(X, Y) = \frac{\Gamma(\frac{d-1}{2}\pm i\nu)}{(4\pi)^{d/2}\Gamma(\frac{d}{2})\Gamma(\pm i\nu)} \, _2F_1\left(\frac{d-1}{2}+i\nu, \frac{d-1}{2}-i\nu;\frac{d}{2}; \frac{1+X\cdot Y}{2} \right)~,
\end{align}
where $X\in\mathbb R^{1, d}$ is the embedding space coordinate of AdS.
The harmonic functions satisfy the completeness and orthogonality conditions
\begin{align}\label{CO_harm}
\int_{\mathbb R} d\nu \,\Omega_\nu(X, Y) = \delta(X, Y)\,, \quad 
\int_Y \Omega_{\nu}(X, Y)\Omega_{\bar\nu}(Y, Z) = \frac{\delta(\nu-\bar\nu)+\delta(\nu+\bar\nu)}{2}\Omega_\nu(X, Z)~,
\end{align}
because they form an eigenbasis of the Laplacian operator of AdS$_d$
\begin{align}\label{eigen}
-\nabla_X^2 \Omega_\nu(X, Y) = -\nabla_Y^2 \Omega_\nu(X, Y) = \left(\frac{(d-1)^2}{4}+\nu^2\right)\Omega_\nu(X, Y)~.
\end{align}

Given an AdS invariant two-point function $G(X, Y)$, we expand it into AdS harmonic functions 
\begin{align}\label{Gr_app}
G(X, Y) = \int_{\mathbb R} d\nu\, \rho_G(\nu)\,\Omega_\nu(X, Y)~.
\end{align}
We call $\rho_G(\nu)$  the spectral density of $G$.
For a free scalar of boundary dimension $\hat\Delta$, the corresponding spectral density is $1/(\nu^2+(\hat\Delta-\frac{d-1}{2})^2)$. The spectral density has some useful properties:
\begin{itemize}
\item The two-point function $G(X, Y)$ defines an operator in AdS, sending $f(X)$ to $\int_Y G(X, Y)f(Y)$. Let $H$ be the inverse of $G$, i.e. $\int_Y G(X, Y)H(Y, Z) = \delta(X, Z)$, then the spectral density of $H$ is simply $1/\rho_G(\nu)$.
\item The functional determinant of $G$ is completely encoded in its spectral density
\begin{align}\label{logdet_general}
\frac{1}{2}\log \det(G) = \frac{{\rm V}_d}{(4\pi)^{d/2}\Gamma(\frac{d}{2})}\int_0^\infty d\nu\, \frac{\Gamma(\frac{d-1}{2}\pm i\nu)}{\Gamma(\pm i\nu)} \log (\rho_G(\nu))~,
\end{align}
where ${\rm V}_d =\pi^{\frac{d-1}{2}}\Gamma(\frac{1-d}{2})$ is the volume of AdS$_d$.
\end{itemize}
One of the main technical tasks in this paper is computing the spectral density given a two-point function $G$
\begin{align}\label{rhoG}
\rho_G(\nu) &= \frac{1}{\Omega_\nu(X, X)}\int_Y G(X, Y)\Omega_\nu(X, Y)\nonumber\\
& = {\rm Vol}(S^{d-1})\int_0^\infty dr\, \sinh^{d-1}(r)\, G(r)\, _2F_1\left(\frac{d-1}{2}+i\nu, \frac{d-1}{2}-i\nu;\frac{d}{2}; \frac{1-\cosh(r)}{2} \right)~,
\end{align}
where we have used the orthogonality condition. The variable $r$ arises naturally as we move $X$ to the origin of AdS and use the global coordinate for $Y$.

\subsection{A class of integrals}\label{app:integrals}
The spectral density of two-point functions that are proportional to a power of $\sinh(r)$ can be computed analytically. The relevant integral is 
\begin{align}
S_\kappa(\nu)&\equiv \int_0^\infty d r\, \sinh^\kappa(r)\, _2F_1\left(\frac{d-1}{2}+ i\nu, \frac{d-1}{2}- i\nu; \frac{d}{2}; \frac{1-\cosh(r)}{2}\right)~.
\end{align}
For the hypergeometric function in $S_\kappa(\nu)$, we use the following identity 
\begin{align}
_2F_1\left(a, b; \frac{a+b+1}{2};z \right) = (1-2z)^{-a} \,_2F_1\left(\frac{a}{2}, \frac{a+1}{2};\frac{a+b+1}{2};\frac{4z(z-1)}{(1-2z)^2} \right)~,
\end{align}
which yields
\begin{align}
  _2F_1\left(\frac{d-1}{2}+i\nu, \frac{d-1}{2}-i\nu;\frac{d}{2}; \frac{1-\cosh(r)}{2} \right) = \frac{1}{\cosh(r)^{\Delta_\nu}}\,_2F_1\left(\frac{\Delta_\nu}{2}, \frac{\Delta_\nu+1}{2}; \frac{d}{2}; \tanh^2(r)\right)~,
 \end{align}
 where $\Delta_\nu \equiv \frac{d-1}{2}+i\nu$. Using the series definition of hypergeometric functions and evaluating the $r$ integral term by term, we find
\begin{align}\label{Sk_result}
S_\kappa(\nu)=\frac{\Gamma(\frac{d}{2})\Gamma(\frac{\kappa+1}{2})}{2\Gamma(\frac{d-\kappa-1}{2})}\hat g_{\frac{d-2\kappa-1}{4},\frac{d+1}{4}}(\nu) \,, \quad \hat g_{a, b}(\nu)=\frac{\Gamma(a\pm i\frac{\nu}{2})}{\Gamma(b\pm i\frac{\nu}{2})}~.
\end{align}

\subsection{Osborn's trick}
Osborn defined a series of transformations to invert a bulk two-point function in BCFT \cite{McAvity:1995zd}. We show here that after mapping the BCFT to AdS, Osborn's trick is equivalent to computing the spectral density of the two-point function. Let us first briefly review his method in the AdS set-up \cite{Giombi:2020rmc}. Let $G(\xi)$ and $H(\xi)$ be a pair of  AdS invariant two point functions, with $\xi = \frac{-X\cdot Y-1}{2}$. The transformations are defined as follows
\begin{align}
g( u ) = \frac{\pi^{\frac{d-1}{2}}}{\Gamma(\frac{d-1}{2})} \int_0^\infty d\xi\, \xi^{\frac{d-3}{2}} G(\xi+u)\,, \quad \, \hat g(k) = \int_{\mathbb R} d\theta\, e^{i k\theta} g(\sinh^2\theta)~,
\end{align}
and similarly for $H$. It was observed in \cite{McAvity:1995zd} that if $H$ is the inverse of $G$, then $\hat g(k)\hat h(k) = 1/4^d$. 
We claim that,  up to normalizations, $\hat g$ is essentially the same as the spectral density of $G$,  defined by \eqref{Gr_app}. 
A crucial tool is the following integral representation of $\Omega_\nu$ 
\begin{align}\label{Krep}
\Omega_\nu (X, Y) =\frac{1}{(2\pi)^{\frac{d+1}{2}}\Gamma(\pm i\nu)}\int_0^\infty \frac{ds}{s}s^{\frac{d-1}{2}} K_{i\nu}(s) e^{s X\cdot Y}~.
\end{align}
This integral yields a Legendre function, whose  hypergeometric function representation is  $\Omega_\nu$.
Combining \eqref{Gr_app} and \eqref{Krep} gives 
\begin{align}
G(\xi ) = \frac{1}{(2\pi)^{\frac{d+1}{2}}}\int_{\mathbb R} d\nu\, \frac{\rho_G(\nu)}{\Gamma(\pm i\nu)} \int_0^\infty \frac{ds}{s}s^{\frac{d-1}{2}} K_{i\nu}(s) e^{-(2\xi+1)s}~,
\end{align}
which is a convenient representation for computing $g(\sinh^2\theta)$
\begin{align}
g(\sinh^2\theta) &= \frac{1}{2^{\frac{d+1}{2}}\pi \Gamma(\frac{d-1}{2})}\int_{\mathbb R} d\nu\, \frac{\rho_G(\nu)}{\Gamma(\pm i\nu)} \int_0^\infty \frac{ds}{s}s^{\frac{d-1}{2}} K_{i\nu}(s) e^{-s\cosh(2\theta)}\int_0^\infty \frac{d\xi }{\xi}\xi^{\frac{d-1}{2}} e^{-2s\xi}\nonumber\\
&=\frac{1}{2^{d}\pi}\int_{\mathbb R} d\nu\, \frac{\rho_G(\nu)}{\Gamma(\pm i\nu)} \int_0^\infty \frac{ds}{s} K_{i\nu}(s) e^{-s\cosh(2\theta)}~.
\end{align}
The Fourier transformation with respect to $\theta$ yields another Bessel function  
\begin{align}
\hat g(k) &= \frac{1}{2^{d-1}\pi}\int_{\mathbb R} d\nu\, \frac{\rho_G(\nu)}{\Gamma(\pm i\nu)} \int_0^\infty \frac{ds}{s} K_{i\nu}(s) \int_0^\infty d\theta \cos(k\theta)e^{-s\cosh(2\theta)}\nonumber\\
&= \frac{1}{2^{d}\pi}\int_{\mathbb R} d\nu\, \frac{\rho_G(\nu)}{\Gamma(\pm i\nu)} \int_0^\infty \frac{ds}{s} K_{i\nu}(s) K_{i k/2}(s)\nonumber\\
&=\frac{1}{2^d}\int_0^\infty ds\, K_{ik/2}(s) \left(\int_0^\infty d\nu\, \frac{2\nu\sinh(\pi\nu)}{\pi^2 s}\rho_G(\nu) K_{i\nu}(s)\right) ~.
\end{align}
Using the Kontorovich-Lebedev transform and its inverse transformation 
\begin{align}
\bar f (\nu) = \int_0^\infty ds\, f(s) K_{i\nu} (s)\,, \quad f(s) = \int_0^\infty d\nu\, \frac{2\nu\sinh(\pi\nu)}{\pi^2 s}\bar f(\nu) K_{i\nu}(s)~,
\end{align}
we can immediately conclude $2^d \hat g(2\nu) = \rho_G(\nu)$. Altogether, Osborn's method of inverting two-point functions is the same as inverting the corresponding spectral density of the two-point function.

\subsection{The extraordinary fixed point of the O$(N)$ model}\label{sec:extraordinary}
 Following  \cite{Giombi:2020rmc}, the symmetry-breaking fixed point can be found by
integrating out the first $N-1$ $\phi^I$ fields. It generates  an action for $\phi^N$ and $\sigma$:
\begin{align}
S = \int d^d x\sqrt{g}\left[\frac{1}{2}(\partial \phi^N)^2+\frac{1}{2}(\phi^N)^2\left(\sigma\!-\! \frac{d(d-2)}{4}\right) \!+\! \frac{N-1}{2}\tr\log\left(-\nabla^2\!+\!\sigma\!-\!\frac{d(d-2)}{4}\right)\right]~.
\end{align}
At large $N$, the saddle point corresponding to the extraordinary  fixed point is
\begin{align}
\sigma_\star = \frac{d(d-2)}{4}~, \qquad (\phi^N_\star)^2 = -\frac{(N-1)\Gamma(d-1)\Gamma(1-\frac{d}{2})}{(4\pi)^{d/2}}~.
\end{align}
Then the $N-1$ transverse fields are massless scalars with boundary dimension $\hat\Delta = d-1$.  They are the tilt operators. We denote their propagator by $G_0$:
\begin{align}
G_0 = \frac{\Gamma(\frac{d}{2})}{(4\pi)^{d/2}(d-1)\xi^{d-1}}\, \F\!\left(d-1, \tfrac{d}{2}; d; -\tfrac{1}{\xi}\right)~.
\end{align}
Consider fluctuations around the saddle point: $\sigma = \sigma_\star +i \delta\sigma$ and $\phi^N = \phi^N_\star + \chi$. The quadratic action of $\chi$ and $\delta\sigma$ is
\begin{align}
S_2=\int d^d x\sqrt{g}\left[\frac{1}{2}(\partial \chi)^2+i\phi^N_\star \chi\,\delta\sigma \right]+\frac{N-1}{4} \int d^d x\sqrt{g_x}\, d^dy\sqrt{g_y}\, G_0(x, y)^2\, \delta\sigma(x)\,\delta\sigma(y)~.
\end{align}
Integrating out the longitudinal mode $\chi$ generates an additional contribution to the $\delta\sigma$ kernel. The resulting quadratic action for $\delta\sigma$ alone is
\begin{align}
\tilde S_2 = \frac{1}{2} \int d^d x\sqrt{g_x}\, d^dy\sqrt{g_y}\, K_0(x, y)\, \delta\sigma(x)\,\delta\sigma(y)~, \quad K_0 = \frac{N-1}{2}G_0^2+(\phi^N_\star)^2 G_0~.
\end{align}

Computing the one-loop free energy now requires the spectral density of $K_0$. The spectral density of $G_0$ is known: $\rho_{G_0}(\nu) = \frac{1}{\nu^2+(\frac{d-1}{2})^2}$. The new task is to find the spectral density of $G_0^2$. A useful trick is to compute $-\nabla^2 G_0^2$ first \cite{McAvity:1995zd}
\begin{align}\label{DG0}
-\nabla_X^2 G_0^2(X, Y) =-\frac{ \Gamma (\frac{d}{2})^2 }{2^{2 d-1}\pi ^{d} (\xi  (\xi +1))^{d-1}} ~.
\end{align}
The function $(\xi(\xi+1))^{1-d}$ also appears in the normal boundary of the GN model, and its spectral density is:
\begin{align}\label{d2G2}
\frac{N-1}{2}(-\nabla^2_X) G^2_0(X, Y) = \int_{\mathbb R}d\nu\, \tilde \rho(\nu)\,\Omega_\nu(X, Y)~, \quad \tilde\rho(\nu) = -\frac{(N-1)\hat g_{\frac{3(d-1)}{4},\frac{d+1}{4}}(\nu)}{2^d\pi^{\frac{d-3}{2}}\sin(\frac{d\pi}{2})\Gamma(\frac{d-1}{2})}~.
\end{align}
Because the harmonic functions are eigenfunctions of the Laplacian operator, cf. \eqref{eigen}, we can straightforwardly invert the action of $-\nabla^2$. The inversion yields the spectral density $ \frac{\tilde\rho(\nu)}{\nu^2+(\frac{d-1}{2})^2}$ for $\frac{N-1}{2}G_0^2$.
Altogether, the order $N^0$ of the free energy is
 \begin{align}
\CF^{(1)}_{\rm ext} =  -\frac{{\rm V}_d}{(4\pi)^{d/2}\Gamma(\frac{d}{2})}\int_0^\infty d\nu\, \frac{\Gamma(\frac{d-1}{2}\pm i\nu)}{\Gamma(\pm i\nu)} \log \Big[(\nu^2+\tfrac{(d-1)^2}{4})\,\hat g_{\frac{d+1}{4}, \frac{3(d-1)}{4}}(\nu)\Big]~.
\end{align}
For $d=3$, simply by comparing the spectral density, we find $\CF^{(1)}_{\rm ext} = \CF^{(1)}_{\rm ord} = \frac{\log(R)}{48}$. It is in agreement with \cite{Krishnan_2023}. To extract the $\epsilon$-expansion, we split $ \CF^{(1)}_{\rm ext} $ into two parts 
$
\CF^{(1)}_{\rm ext}  = -F_{\hat\Delta = d-1} +  F^{(1)}_{\rm nor}$,
where $F^{(1)}_{\rm nor}$ is the one-loop free energy of the GN normal phase, given by \eqref{FB1order1}.
$F_{\hat\Delta = d-1} $ denotes the free energy of a free scalar of boundary dimension $d-1$, whose $\epsilon$-expansion  was calculated in \cite{Giombi:2025pxx}
\begin{align}
-F_{\hat\Delta = d-1} =\frac{29}{180\epsilon}+0.0057916+0.265129\epsilon+O\left(\epsilon^2\right)\,.
\end{align}
Combining the two contributions gives
\begin{align}
\CF^{(1)}_{\rm ext}=-\frac{4}{3\epsilon}+2.66403-4.29451\epsilon+O\left(\epsilon^2\right)~.
\end{align}

\section{Computation of $F_{\rm free}$ }\label{FreeEnergyMasslessFermions}
In this appendix, we evaluate the free energy of a massless Dirac fermions in AdS$_{d=4-\epsilon}$ to order $\epsilon$. The spectrum representation of the free energy reads
\begin{align}\label{Ffreespec}
F_{\rm free} =&-\frac{{\rm V}_dc_d}{(4\pi)^{\frac{d}{2}}\Gamma\left(\frac{d}{2}\right)}\int \limits_{0}^{+\infty}d\lambda \mathcal{P}_{\epsilon}(\lambda)\log \left(\lambda^2\right)\,, \qquad \mathcal{P}_{\epsilon}(\lambda)=\frac{\Gamma\left(\frac{d}{2}+i\lambda\right)\Gamma\left(\frac{d}{2}-i\lambda\right)}{\Gamma\left(\frac{1}{2}+i\lambda\right)\Gamma\left(\frac{1}{2}-i\lambda\right)}\,.
\end{align}
We compute the $\epsilon$-expansion using the subtraction method. Based on the large $\lambda$ asymptotic of $\CP_\epsilon(\lambda)$ derived using
Stirling's formula, we choose the following subtraction function:
\begin{align}
    &\hat{\mathcal{P}}_{\epsilon}(\lambda)=\left(\lambda^2+4\right)^{\frac{d-1}{2}}\Bigg(1\!+\!\frac{(d-8) (d-1) (d+6)}{24 \left(\lambda ^2+4\right)}\!+\!\frac{(d-3) (d-1) \left(5 d^4\!-\!28 d^3\!-\!452 d^2\!+\!976 d\!+\!11520\right)}{5760 \left(\lambda ^2+4\right)^2}\nonumber\\
    &+\frac{(d-5) (d-3) (d-1) \left(35 d^6-378 d^5-3988 d^4+28200 d^3+212384 d^2-500736 d-3870720\right)}{2903040 \left(\lambda ^2+4\right)^3}\Bigg)\,.
\end{align}
We then decompose the integral as
\begin{equation}\label{CPdec}
    \begin{split}
        \int \limits_{0}^{+\infty}d\lambda \mathcal{P}_{\epsilon}(\lambda)\log \left(\lambda^2\right)=&\int \limits_{0}^{+\infty}d\lambda \hat{\mathcal{P}}_{\epsilon}(\lambda)\log \left(\lambda^2\right)+\int \limits_{0}^{+\infty}d\lambda \left(\mathcal{P}_{0}(\lambda)-\hat{\mathcal{P}}_{0}(\lambda)\right)\log \left(\lambda^2\right)\\
    &+\epsilon\int \limits_{0}^{+\infty}d\lambda \left(\mathcal{P}'_{0}(\lambda)-\hat{\mathcal{P}}'_{0}(\lambda)\right)\log \left(\lambda^2\right)+O(\epsilon^2)\,.
    \end{split}
\end{equation}
The first integral can be evaluated using
\begin{align}
    \int \limits_{0}^{+\infty}d\lambda \frac{\log(\lambda^2)}{\left(\lambda^2+4\right)^a}=-\sqrt{\pi }\frac{ \Gamma \left(a-\frac{1}{2}\right)  }{2^{2a}\Gamma (a)}H_{a-\frac{3}{2}}\,,
\end{align}
and the second integral can also be calculated analytically:
\begin{gather}
    \int \limits_{0}^{+\infty}d\lambda \left(\mathcal{P}_{0}(\lambda)-\hat{\mathcal{P}}_{0}(\lambda)\right)\log \left(\lambda^2\right)
    =\frac{1}{360} \left(161-720 \log (A)+\frac{6 \zeta '(4)}{\zeta (4)}-6 \gamma_E -6 \log (2 \pi )\right)\,.
\end{gather}
Numerically evaluating the last integral in \eqref{CPdec} yields:
\begin{gather}
    \epsilon\int \limits_{0}^{+\infty}d\lambda \left(\mathcal{P}'_{0}(\lambda)-\hat{\mathcal{P}}'_{0}(\lambda)\right)\log \left(\lambda^2\right)=-0.15097\epsilon\,.
\end{gather}
Putting everything together, we obtain
\begin{gather}
    F_{\rm free}=\frac{11}{180 \epsilon }-\frac{41}{432}+\frac{2 \log (A)}{3}-\frac{\zeta '(4)}{2 \pi ^4}-\frac{\gamma_E }{18}+\frac{\log (2 \pi )}{180} +0.04593 \epsilon
\end{gather}
which matches \eqref{Ffree}.

\section{Details of the two-loop computation of $\CF_{\textit{2-loop}}^{(6)}$}\label{App:SubtractionFunction}
In this appendix, we provide more details about computing $\CF_{\textit{2-loop}}^{(6)}$, c.f. \eqref{F6diagram}.  As the first step, it can be rewritten in the form
\begin{gather}
    \CF_{\textit{2-loop}}^{(6)}=\frac{g^2_{1}N}{2}{\rm V}_d{\rm Vol}(S^{d-1})\int \limits_{1}^{+\infty} d u k_{\epsilon}(u,\mu)\,, \quad
   k_{\epsilon}(u,\mu)=\frac{G_t(u)\tr{\left(G^{+}_{\mu}(x,y)G^{+}_{\mu}(y,x)\right)}}{\left(u^2-1\right)^{1-\frac{d}{2}} }\,.
\end{gather}
where $u=2\xi+1$ and $\mu=\frac{g_{1}}{\sqrt{g_{2}}}\sqrt{\frac{3d(d-2)}{2}}$. Using the identity 
\begin{gather}
    \tr\left\{\left(-\frac{ \gamma_0\slashed{\bar x}_{12}  }{\sqrt{z_1 z_2 } }\frac{\alpha(\xi)}{\sqrt{\xi+1}}+\frac{ \slashed{x}_{12}  }{\sqrt{z_1 z_2 } }\frac{\beta(\xi)}{\sqrt{\xi}}\right)\left(-\frac{ \gamma_0\slashed{\bar x}_{21}    }{\sqrt{z_1 z_2 } }\frac{\alpha(\xi)}{\sqrt{\xi+1}}+\frac{ \slashed{x}_{21} }{\sqrt{z_1 z_2 } }\frac{\beta(\xi)}{\sqrt{\xi}}\right)\right\}
    =4^2\left(\alpha^2-\beta^2\right)\,,
\end{gather}
together with standard transformation identities for hypergeometric functions, we arrive at
\begin{align}
    &\tr{\left(G^{+}_{\mu}(x,y)G^{+}_{\mu}(y,x)\right)}=\frac{\Gamma \left(\frac{d}{2}+\mu \right)^2}{u ^{d+2 \mu +6} \pi ^{d-1} 2^{d+2 \mu +2} (d-2 \mu )^2 \Gamma \left(\mu +\frac{3}{2}\right)^2}\nonumber\\
    &\times \bigg(4u ^4 \mathbb{F}^2_{\mu;0}\left(\frac{1}{u^2}\right) (2 \mu +1)^2  \left(u ^2 (3 d+2 \mu ) (d-2 \mu )+(d+2 \mu )^2\right)\\
    &+\left(u ^2-1\right)^2\mathbb{F}^2_{\mu;1}\left(\frac{1}{u^2}\right)  (d+2 \mu )^2 (d+2 \mu +2)^2\nonumber\\
    &-4u ^2 \left(u ^2-1\right) \mathbb{F}_{\mu;0}\left(\frac{1}{u^2}\right) \mathbb{F}_{\mu;1}\left(\frac{1}{u^2}\right) (2 \mu +1)  (d+2 \mu ) (d+2 \mu +2) \left(u ^2 (d-2 \mu )+d+2 \mu \right)\bigg)\nonumber\,,
\end{align}
where
\begin{equation}\label{bF}
    \begin{split}
        &\mathbb{F}_{\mu;0}\left(\frac{1}{u^2}\right)={}_2F_1\left(\frac{d}{4}+\frac{\mu}{2} ,\frac{d+2}{4}+\frac{\mu}{2} ;\mu+\frac{1}{2};\frac{1}{u^2}\right)\,,\\
    &\mathbb{F}_{\mu;1}\left(\frac{1}{u^2}\right)={}_2F_1\left(\frac{d+4}{4}+\frac{\mu}{2} ,\frac{d+6}{4}+\frac{\mu}{2} ;\mu+\frac{3}{2};\frac{1}{u^2}\right)\,.
    \end{split}
\end{equation}
The function $k_{\epsilon}(u,\mu)$ is singular in the coincident-point limit, corresponding to $u \to 1$. In order to isolate the divergent terms systematically, it is convenient to use the following transformation of the hypergeometric function:
\begin{equation} \label{2F1Transformation_x_to_1}
    \begin{split}
        _2F_1\left(a,b;c;x\right)&=\frac{\Gamma(c)\Gamma(c-a-b)}{\Gamma(c-a)\Gamma(c-b)} {}_2F_1\left(a,b,a+b-c+1;1-x\right){}\\
   & +(1-x)^{c-a-b}\frac{\Gamma(c)\Gamma(a+b-c)}{\Gamma(a)\Gamma(b)} {}_2F_1\left(c-a,c-b,c-a-b+1;1-x\right){}\,,
    \end{split}
\end{equation}
which is particularly well suited for expanding the integrand near $x=1$. We then construct a subtraction function $\hat{k}_\epsilon(u,\mu)$ such that the integral $\int\limits_{1}^{+\infty} du (k_\epsilon(u,\mu)-\hat{k}_\epsilon(u,\mu))$ is convergent in a neighborhood of $\epsilon=0$, while $\int\limits_{1}^{+\infty} du \hat{k}_\epsilon(u,\mu)$ can be evaluated analytically in dimensional regularization using
\begin{gather}
    \int \limits_{1}^{+\infty}du \frac{1}{(u^2-1)^{a}u^{b}}=\frac{\Gamma (1-a) \Gamma \left(\frac{2a+b-1}{2}\right)}{2 \Gamma \left(\frac{1+b}{2}\right)}\,.
\end{gather} 
By applying transformation \eqref{2F1Transformation_x_to_1} to the scalar propagator \eqref{scalar_t_GreenFunction} and \eqref{bF},  we find the following subtraction function that does this job
\begin{gather}
    \hat{k}_\epsilon(u,\mu)=\frac{A_{d-1}(u)}{(u^2-1)^{d-1}}+\frac{A_{d-2}(u)}{(u^2-1)^{d-2}}+\frac{A_{d-3}(u)}{(u^2-1)^{d-3}}
    +\frac{A_{\frac{d}{2}}(u)}{(u^2-1)^{\frac{d}{2}}}+\frac{A_{\frac{d}{2}-1}(u)}{(u^2-1)^{\frac{d}{2}-1}}\,,
    \label{kephatCoefficients}
\end{gather}
where the coefficients $A_{i}(u)$ which are regular in the limit $u \to 1$ are given by
\begin{align}
A_{d-1}(u)&=-\frac{1}{u ^{\hat{\Delta}_t +2 \mu +4-2 d}}\frac{1}{4\pi^{\frac{3 d}{2}}}  \Gamma \left(\frac{d}{2}-1\right) \Gamma \left(\frac{d}{2}\right)^2\,,\\
 A_{d-2}(u)&=\frac{1}{u ^{\hat{\Delta}_t +2 \mu +6-2 d}}\frac{\Gamma \left(\frac{d}{2}-2\right) \Gamma \left(\frac{d}{2}-1\right)^2}{128 \pi ^{\frac{3 d}{2}}}\bigg(4 (d-4) (d-1) \mu ^2-4 (d-4) (d-2)^2 \mu\nonumber\\
    &+(d-2)^2 ( (4 d-2 \hat{\Delta}_t -19)d+8 (\hat{\Delta}_t+3))\bigg)\,,\nonumber
    \label{kephatCoefficientsExplicit1}
\end{align}
and 
\begin{align}
A_{d-3}(u)&=\frac{-1}{u^{\hat{\Delta}_t +2 \mu +8-2 d}}\frac{\Gamma \left(\frac{d}{2}-3\right) \Gamma \left(\frac{d}{2}-1\right)^2}{2048 \pi ^{\frac{3 d}{2}}}\bigg(16 (d-6) (d-1) \mu ^4-32 (d-6) (d-4) (d-1) \mu ^3\nonumber\\
    &+16 (d-6)((4-d) (d-1) \hat{\Delta}_t +3 d ((d-7) d+16)-33)\mu ^2\nonumber\\
    &-8 (d-6) (d-2)^2(d (4 d-2 \hat{\Delta}_t -21)+8 (\hat{\Delta}_t +4))\mu\nonumber\\
    &+(d-2)^2 (d (d (d (18 d-12 \hat{\Delta}_t -221)+136 \hat{\Delta}_t +990)-8 (62 \hat{\Delta}_t +239))+48 (14 \hat{\Delta}_t +29))\bigg),\nonumber\\
A_{\frac{d}{2}}(u)&=-\frac{1}{u ^{\hat{\Delta}_t +2 \mu +2-d}}\frac{   \Gamma \left(\frac{d}{2}\right) \Gamma (\hat{\Delta}_t )}{2^{d}\pi ^{\frac{3 d}{2}-1}\Gamma (\hat{\Delta}_t-d +2)}\csc \left(\frac{\pi  d}{2}\right)\,,\\
A_{\frac{d}{2}-1}(u)&=\frac{1}{u ^{\hat{\Delta}_t +2 \mu +4-d}}\frac{ \csc \left(\frac{\pi  d}{2}\right) \Gamma \left(\frac{d}{2}-1\right)}{2^{d+3} \pi ^{\frac{3 d}{2}-1}(d-2)}\Bigg(\frac{16 (d-1)   \Gamma \left(\frac{d}{2}+\mu \right)}{d \Gamma \left(1-\frac{d}{2}+\mu\right)}\mu\nonumber\\
    &+\frac{\Gamma (\hat{\Delta}_t )}{\Gamma (\hat{\Delta}_t-d +2)}\left(4 (d-1) \mu ^2-4 (d-2)^2 \mu -(d-2)^2 (2 \hat{\Delta}_t +1)\right)\Bigg)\,.\nonumber
    \label{kephatCoefficientsExplicit2}
\end{align}
Then, following the procedure described above we get result in \eqref{F6diagramFinal}.

\bibliography{draft}
\bibliographystyle{utphys}

\end{document}